\documentclass[superscriptaddress,amssymb,twocolumn,aps]{revtex4}

\usepackage{graphicx,dcolumn,bm,amsmath,amssymb,color,bm}

\usepackage[version=3]{mhchem} 

\newcommand{\Will}[1]{\textcolor{black}{#1}}

\definecolor{forestgreen}{rgb}{0.13, 0.55, 0.13}
\newcommand{\Rik}[1]{\textcolor{black}{#1}}

\begin{document}

\title{Complex Smectic Phases in Polymer Grafted Shape-Uniform Cellulose Nano-Crystals}

\author{William S. Fall}
\email{william.fall@universite-paris-saclay.fr}
\affiliation{Laboratoire de Physique des Solides - UMR 8502,
CNRS, Universit\'e Paris-Saclay, 91405 Orsay, France}

\author{Henricus H. Wensink}
\email{rik.wensink@universite-paris-saclay.fr}
\affiliation{Laboratoire de Physique des Solides - UMR 8502,
CNRS, Universit\'e Paris-Saclay, 91405 Orsay, France}

\begin{abstract}
The effect of short chain grafting on the liquid crystalline (LC) ordering of nano-crystals is investigated using molecular dynamics simulations of a coarse-grained grafted nano-rod model. Monodisperse nano-rods, with aspect ratios typical of cellulose nano-crystals (CNCs) are grafted randomly with oligomers at different grafting densities. LC ordering depends non-trivially on grafting density as the effective nano-rod shape and softness is modified.  Ungrafted rods exhibit Nematic and tilted Smectic-C phases. At 25\% grafting, the addition of a few side chains entirely supresses Smectic order and instead a persistent Nematic phase is favoured. Intermediary grafting, around 50\%, results in a Nematic and the reappearance of tilted Smectic-F phases. Heavier grafting facilitates direct transitions to either Smectic-I with extreme tilt (75\%) or an un-tilted Smectic-B (100\%).  Such behaviour falls outside of current hard or soft-rod descriptions of phase-transitions in rod-like LC systems and points to undiscovered LC behaviour in both shape-purified grafted/un-grafted CNCs.
\end{abstract}

\date{\today}

\maketitle

\section{Introduction}

Cellulose is the most abundant polymeric raw material on the planet and is found almost exclusively in the cell-walls of plants \cite{klemm2005cellulose,habibi2010cellulose}. It is environmentally friendly and sustainable to produce, on industrial scales and can be processed into nano-structured materials \cite{moon2011cellulose}. Typically cellulose chains may be extracted by disassembling the structure of plant cells (trees) and turned into nano-crystalline bundles by acid hydrolosis \cite{fujisawa2022structure}. The resulting nano-crystals vary in size and shape with aspect ratios ranging between 10 and 100, depending on the processing conditions \cite{lagerwall2014cellulose,korolovych2018cellulose}. Its structure is highly crystalline, which means it is incredibly strong, with a greater axial elastic modulus than that of Kevlar and its surface is covered with reactive OH groups which facilitates chemical grafting. This makes it one of the most prominent ‘green’ materials of modern times which can be used in bio-based functional materials \cite{isogai2013wood,trache2020nanocellulose}. Examples include polymer films and eco packaging, flexible displays, reinforcing agents for polymers and actuators, biomedical implants, pharmaceuticals, textiles, templates for electronic components, membranes, batteries, super-capacitors and many others \cite{moon2011cellulose}. In many of the above applications, the alignment of CNCs in suspension is essential for device performance particularly in displays or when used as reinforcing agents. 

The LC behaviour of neat CNCs in solution is relatively well known experimentally, with Nematic and chiral Nematic phases being observed, driven by the twisted nature of the CNCs \cite{parton2022chiral} as well as Isotropic-Nematic phase coexistence driven by their elongated shape and inherent polydispersity \cite{lagerwall2014cellulose,da2022cellulose}. The precise twisted nature of CNCs  \cite{ogawa2019electron} and the transfer of chirality across length-scales have been subject of recent experimental  \cite{honorato2017cholesteric,parton2022chiral,gonccalves2021chirality,fittolani2022bottom} and modelling studies \cite{wensink2019effect,chiappini2022modeling,sewring2023effect}. Another level of complexity can be achieved by adding polymers to suspensions to CNCs. The polymers conformational degrees of freedom lead to effective interactions between CNCs that are drastically different from those between naked CNCs \cite{sun2022polymer}. 
If the polymers are strictly non-adsorbing the main effect incurred is the so-called  depletion attraction, which originates from polymers being depleted away from the interior space between two colloids at close proximity and generates an osmotic imbalance pushing the colloids together.  A considerable body of literature has been devoted to exploring this phenomenon and its consequences for the phase behavior for a wide range of colloidal shapes including rods \cite{lekkerkerker2011depletion}. 

When polymers adsorb onto the colloid surface the depletion effect is compounded with reversible or irreversible polymer-substrate adhesion and the impact of the presence of the polymers on the self-assembly is far less clear \cite{gregory1978effects}.  Theoretical studies usually resort to extreme coarse-grained models in terms of a polymer-mediated effective shape or compressible colloids to gauge the impact on phase behavior \cite{shundyak2006theory,dennison2011phase}. Integral equation theories have been applied to understand the relation between colloid shape and property of polymer-nanocomposites involving sparsely grafted rods \cite{gollanapalli2017dispersion2} and other non-spherical colloids \cite{hall2010structure}. Self-consistent field computations can be employed to construct effective interactions between nano-rods immersed in polymer solutions and explore their phase behavior \cite{surve2007dispersion,ma2013binary}. Effective-shape ``soft rod" models \cite{grelet2016soft,bolhuis1997numerical,cuetos2015liquid} are  useful concepts to explain certain experimental trends in the LC phase behavior of colloids \cite{el2020destabilization} but usually do not give a correct rendering of the subtle role of the polymer degrees of freedom that would follow from an explicit-monomer representation of the adsorbed polymers.       

In recent years, surface modification and polymer grafting of CNCs has gained considerable traction in experiment as the polymers can be used to fine-tune their surface properties and endow the particles with a variety of functionalities \cite{eyley2014surface,wohlhauser2018grafting, kedzior2019recent,zhang2021grafting}. Currently, CNCs grafted with stimuli-responsive polymers are being tested for their actuation properties for use in artificial muscles. A crucial element of this process is the alignment of the nano-rods within the polymer matrix to facilitate anisotropic expansion and increase the efficiency of working devices \cite{ianiro2023computational}. However, to our knowledge no systematic studies into the LC behaviour of polymer grafted CNCs have been performed to date. Whilst simulations of nano-rod-polymer mixtures are now relatively widespread \cite{li2023revealing,toepperwein2011influence,afrasiabian2023dispersion,lu2021molecular,starr2002molecular,savenko2006phase,hu2013depletion,hore2014functional,gao2014molecular,sankar2015dispersion,milchev2020entropic,fall2021canonical,erigi2021phase,gooneie2017length} very few simulations of polymer grafted nano-rods exist and it is only very recently that computational attempts have been made to understand their self-assembly. These studies have been so far limited to single molecules \cite{fujisawa2022structure,joshi2022coarse}, disordered thin films \cite{li2018molecular} in which polymer grafting is noted to disrupt the LC ordering and bulk studies into mechanical properties \cite{shen2018mechanical}, dispersion effects \cite{shen2017insight,chen2018design}, graft locations \cite{li2020effects} and extremely low grafting densities where single polymer chains are attached to the rod tip which are computationally less demanding and allow for their LC behavior to be explored \cite{horsch2005self,wilson2009computer}. In none of these studies however is LC like behaviour observed, likely due to low effective aspect ratios of short rods grafted with long polymers, patchy grafting or improper crystallisation protocols used to grow LC phases. 
Despite the upsurge of experimental work on polymer-grafted nano-rods of various chemical origin  \cite{huang2016well,meuer2008liquid,zorn2008liquid,liu2018filamentous} over the past decades, a systematic computational exploration of the LC behaviour of nano-rods at {\em variable polymer grafting } remains elusive to date.

Here a first step is made in this direction by developing a simple coarse-grained bead model for CNCs which facilitates access to the length scales typical of LC phases in CNCs. Nano-rods are considered to be uniform in length $l_{\mathrm{c}}$ and width $w_{\mathrm{c}}$ and uncharged for simplicity. Our CNCs are distinctly non-cylindrical with anisotropic diamond shaped cross-sections and are slightly flexible with an aspect ratio ($l_{\mathrm{c}}/w_{\mathrm{c}}\approx15$) which reflects their typical average shape \cite{korolovych2018cellulose,lagerwall2014cellulose}. The rods are then grafted with short oligomers of relative length $l_{\mathrm{p}}/l_{\mathrm{c}}=0.13$ at a number of grafting densities ranging from zero to maximum grafting; where every CNC surface site has a polymer attached to it. Interactions between the CNC and polymer beads are purely repulsive and  phase transformations are driven by entropy alone with temperature playing no role in the phase diagram \cite{frenkel1999entropy}. Although naked CNCs are strictly chiral, the propagation of chirality from the microscopic to the LC meso-scale is not discussed since chiral forces are weak and are likely strongly screened by the polymer coating, in particular at elevated grafting densities.  


In the proceeding sections, the coarse-grained model and parameters are first introduced followed by the five separately grafted systems and the strategy for distributing the side chains. The protocol to crystallize these systems into LC phases, is then discussed, including the equilibration procedure and different compression stages. Nematic and Smectic order parameters are then defined and the observed LC ordering is interpreted in terms of each to build the phase diagram. After which, the structure of the different phases is characterized in detail by monitoring 2d and 3d structure factors, the Smectic layer tilt angle and layer spacing as a function of concentration. Dynamics in the  Smectic phases is briefly touched upon and the Van Hove correlation function introduced to assess diffusive layer hopping in the various Smectic symmetries discussed. Finally in the concluding section the importance of these findings is highlighted and related to polydisperse systems typical of real experiments, where undiscovered LC phases are postulated. 

\section{Model and Methods}

\begin{figure}
\includegraphics[width=\columnwidth]{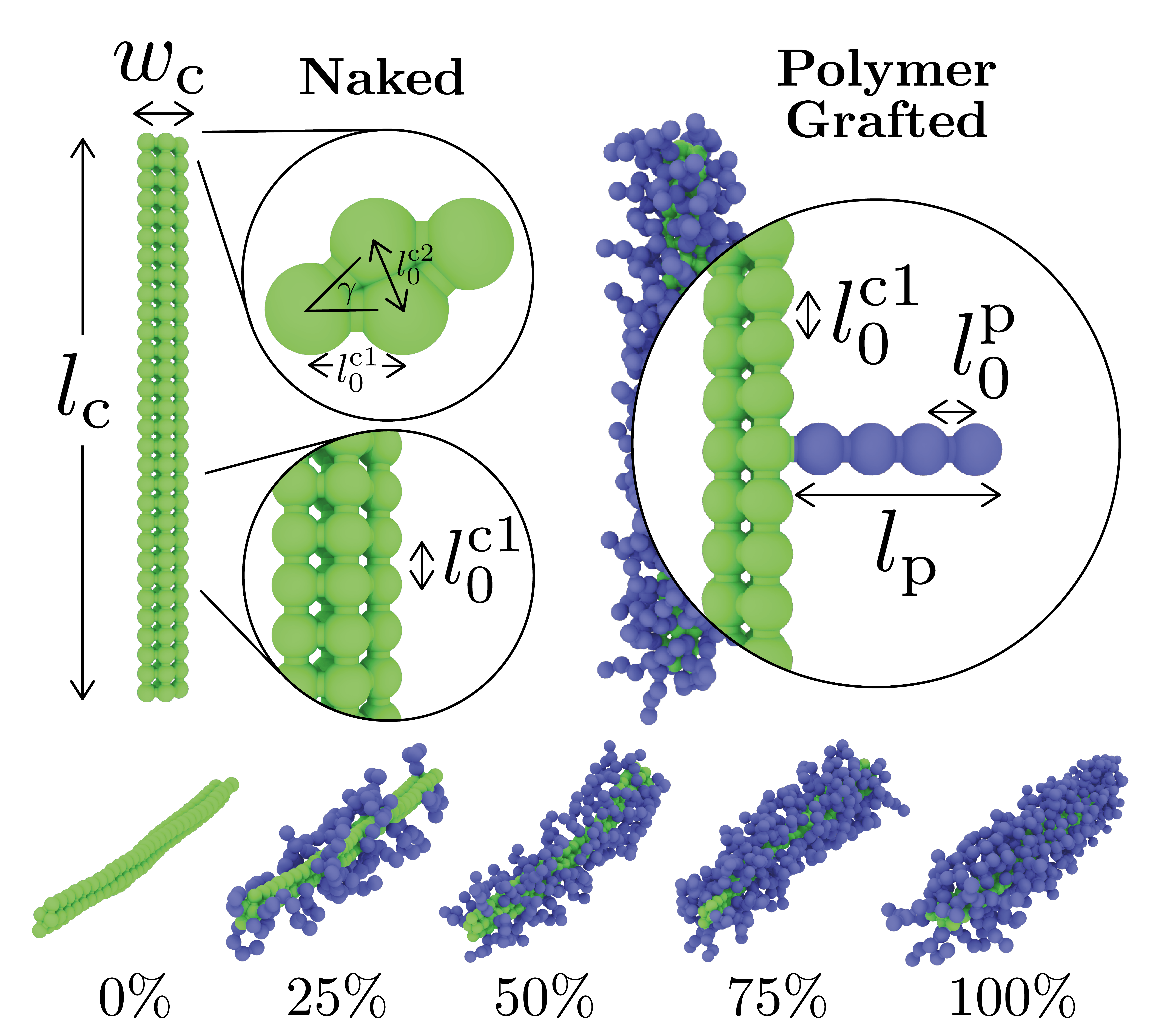}
\caption{\label{fig:model} Colloidal nano-rod model (green) with width $w_{\mathrm{c}}=2\sigma$ and length $l_{\mathrm{c}}=30\sigma$. Nano-rod (green) with grafted linear polymers (purple) of length $l_{\mathrm{p}}=4\sigma$. The bottom 5 grafted nano-rods show the different grafting densities $\rho_{\mathrm{g}}=0,0.25,0.5,0.75$ and 1.0 from left to right respectively and the gradual change in shape as each rod is grafted more heavily.} 
\end{figure}

Standard molecular dynamics is employed, along with a generic modelling strategy designed to study the self-organisation of polymer grafted cellulose nano-crystals (CNCs). Colloidal rods represent the CNCs and are modelled as squashed prism-like point meshes of width $w_{c}$ and length $l_{c}$, where the point mesh of the cross section lies on a 2d rhombic lattice with primitive translation vectors $a=b=\sigma=1.0$nm and an internal angle $\gamma=45^\circ$. A cartoon of the naked colloid is shown in Fig. \ref{fig:model} for clarity.  The point mesh is held together by harmonic bonds of the form 
\begin{eqnarray}
U_{\mathrm{bond}}(l)=\frac{1}{2}k_{\mathrm{bond}}(l-l_{0})^{2}
\label{e:bond}
\end{eqnarray}
where $U_{\mathrm{bond}}(l)$ is the potential energy change associated with deforming the bond with length $l$ from its equilibrium separation $l_{0}$ and $k_{\mathrm{bond}}$ is the spring constant. The longer bonds of the lattice, connecting the outside edges are set to $l_{0}^\mathrm{(c1)}=\sigma$ and the shorter bonds to $l_{0}^\mathrm{(c2)}=0.765~\sigma$, with spring constants $k_{\mathrm{bond}}^\mathrm{(c1)}=k_{\mathrm{bond}}^\mathrm{(c2)}=1000~k_\mathrm{B}T/\sigma^{2}$.  Even though the cross-section of CNCs was found to be strongly disperse in size and shape \cite{rosen2020cross} it appears to be unambiguously non-circular and the effective shape of CNCs thus deviates from a simple uniaxial cylinder.  Our model captures the essential lath-shaped anisotropy of the CNCs using a minimum number of beads to keep our simulations computationally manageable.\cite{rosen2020cross}

To ensure a rigid colloid, additional harmonic angular potentials are enforced between every 3 beads along the colloid which are defined as follows
\begin{eqnarray}
U_{\mathrm{angle}}(\theta)=\frac{1}{2}k_{\mathrm{angle}}(\theta-\theta_{0})^{2}
\label{e:angle}
\end{eqnarray}
where $U_{\mathrm{angle}}(\theta)$ is the potential energy change associated with deforming the angle away from its equilibrium angle, $\theta_{0}^\mathrm{(c)}=180^\circ$ and $k_{\mathrm{angle}}^\mathrm{(c)}=120~k_\mathrm{B}T/\mathrm{rad}^{2}$ is the spring constant.  Note superscripts $\mathrm{(c)}$ and $\mathrm{(p)}$ distinguish between colloid and polymer bonded interaction parameters respectively. In line with experimental observations, the model CNCs are not perfectly rigid but exercise weak conformational fluctuations such that their persistence length strongly exceeds the contour length \cite{usov2015understanding} (see Supplementary Information). Although the CNCs are allowed to slightly twist along their long axis, there is no preferred handedness and the interactions between the colloids are strictly non-chiral.

Non-bonded interactions between colloidal rods take place via a standard WCA potential which is based on the 12-6 Lennard-Jones form 
\begin{eqnarray}
U_{\mathrm{LJ_{12-6}}}=4\epsilon\bigg[\Big(\frac{\sigma}{r}\Big)^{12}-\Big(\frac{\sigma}{r}\Big)^{6}\bigg],\ \ \ r\leq r_{\mathrm{c}}
\label{e:nonbond}
\end{eqnarray}
where $\epsilon=k_\mathrm{B}T=1.0$ denotes the depth of the potential well in terms of the thermal energy with temperature $T$ and Boltzmann's constant $k_\mathrm{B}$, $\sigma$ the zero-crossing (bead size), $r$ the inter-bead separation and $r_{\mathrm{c}}$ the cutoff distance. The potential is cut and shifted to zero at the minimum $r_{\mathrm{c}}=2^{\frac{1}{6}}\sigma$ ensuring only repulsive interactions such that the polymer chains freely explore the surrounding space. The interactions between colloid  and polymer beads are equal such that $\epsilon_{\mathrm{cc}}=\epsilon_{\mathrm{pp}}=\epsilon_{\mathrm{cp}}$ and the mass ratio of different species  $m_{\mathrm{c}}/m_{\mathrm{p}}=1.5$.

Polymers are modelled using simple bead-spring chains, each having the same size $\sigma$ as the beads comprising the colloid but with a reduced mass. The bonded interactions take place via the same potential defined in Eqn \ref{e:bond} where $l_{0}^\mathrm{(p)}=\sigma$, with spring constant $k_{\mathrm{bond}}^\mathrm{(p)}=400~k_\mathrm{B}T/\sigma^{2}$. 
The chain length of the polymer is kept fixed at $l_{\mathrm{p}}=4$ and 5 different grafting densities are considered $\rho_{\mathrm{g}}=0,0.25,0.5,0.75$ and 1, where $\rho_{\mathrm{g}}$ is defined as the ratio of occupied to available surface sites, see Tbl. \ref{tbl:systems}. The grafting sites on the surface of each colloid are chosen at random and no two colloids have the same grafting pattern introducing a source of polydispersity. 

\begin{table}
\caption{\label{tbl:systems} Parameters of all systems considered, $\rho_{\mathrm{g}}$, $M_{\mathrm{c}}$, $N_{\mathrm{c}}$ and $N$ denote the ratio of occupied to available sites or grafting density, the number of beads per colloid, number of colloids and total number of particles respectively in the simulation. Note the system $\rho_{\mathrm{g}}=0.5$ contains 120 more beads to ensure no half particles.}

\begin{tabular}{c|c|c|c|c|}
 $\rho_{\mathrm{g}}$ & $M_{\mathrm{c}}$ & $N_{\mathrm{c}}$ & $N$ \\
\hline
0 & 120 & 2000 &  240,000 \\
0.25 & 240 & 1000 & 240,000 \\
0.5 & 360 & 667 & 240,120 \\
0.75 & 480 & 500 & 240,000 \\
1 & 600 & 400 & 240,000 \\
\end{tabular}
\end{table}

All colloid systems comprised $N=240,000$ beads, with the exception of the system corresponding to $\rho_{\mathrm{g}}=0.5$ which contains 120 more beads to ensure no half molecules, see Tbl. \ref{tbl:systems}. Throughout, $\phi=N\sigma^{3}/V$ and $\phi_{\mathrm{rods}}=N_{\mathrm{CNC}}\sigma^{3}/V$ denote the total \Rik{bead} and rod \Will{ bead concentrations}, where $N$ is the total number of beads in the system, $N_{\mathrm{CNC}}$ is the total number of CNC backbone beads and $V$ is the box volume. These \Rik{dimensionless concentrations} are a measure of the total effective volume fractions of all beads, $\phi$ and the effective volume fractions of only the beads comprising the backbone rods respectively, $\phi_{\mathrm{rods}}$ \footnote{\Rik{It should be emphasized that the packing fraction is a somewhat ambiguous quantity as the colloids and polymers are non-convex objects and the bead interactions are not strictly hard.  Given that our beads are spherical and accounting for the cusps and interstitial spaces of the colloidal nano-rod (see Figure \ref{fig:model}) one is lead to define $\phi = (\pi \sigma^{3}/6)N/V $ as the natural packing fraction \cite{williamson1995excluded}. However, the above definition is believed to be more physically intuitive as it is based on an effective particle volume defined by the envelope or convex hull of the CNC and the polymer which is a more relevant  measure for describing entropy-driven ordering phenomena.  } }, \Rik{so that concentration and volume fraction will be used interchangeably throughout this paper}.   The initial system was prepared by inserting colloids into a box randomly with a homemade code ensuring no overlaps to a specified filling fraction, in this case a concentration of $\phi=0.1$. Equilibration and production runs are then performed using the Large-Scale Atomic/Molecular Massively Parallel Simulator (LAMMPS) \cite{plimpton1995fast,thompson2022lammps}. A short equilibration is performed in the canonical NVT ensemble to obtain the equilibrium pressure before switching to the NPT ensemble for production runs via a Langevin thermostat, with coupling constant $\mathit{\Gamma}=2.0$ ($1/\tau$) and an isotropic Berendsen barostat with $P_{\mathrm{damp}}=100.0$ ($\tau$). The integration timestep 0.005~$\tau$ is used, where the LJ-time unit $\tau = \sqrt{m\sigma^2/(k_\mathrm{B}T)} $, temperature is held fixed at $T=1.0~(k_\mathrm{B}/\epsilon$). The system is run for long times at the equilibrium pressure in order to obtain the Isotropic phase. 

The persistence length of naked CNCs can be approximated based on the bond angle correlation as defined in Equation S1, both along and transverse to the main rod orientation (see Supplementary Information). Taking the isotropic system with $\phi =0.1$ and performing a  fit based on Equation S1 we estimate $L_{P}^{(\parallel)} \sim 700 l_{c}$  and $L_{P}^{(\perp)} \sim 50 l_{c}$ which demonstrates that the CNCs behave as near-rigid rods. However, weak but non-negligible conformational fluctuations, particularly those related to a non-chiral twisting of backbone, impart an additional source of entropy onto our systems and play a subtle role in the phase behavior. As expected, both persistence lengths grow with \Will{concentration} as the rods tend to stiffen up in crowded conditions, see Fig. S2.

After production runs are complete, the pressure is continuously and slowly ramped at a constant rate of $10^{-8}$ ($\epsilon/\sigma^{3}/\tau$)
until liquid crystalline phases with marked orientational order are obtained. At which point, the x,y and z directions are decoupled and an anisotropic barostat is introduced. Switching at a later stage prevents the box dimensions drifting too far away from a cube at early times, whilst allowing for some anisotropy, preventing unwanted stresses building up in the system and mostly eliminating box size effects from forcibly imposed square symmetry. Once the concentration $\phi>0.35$, a series of different compression rates are used to push deeper into the phase diagram. Beginning with $2\times10^{-7}$  ($\epsilon/\sigma^{3}/\tau$) until $\phi>0.6$ after which we switch to $2\times10^{-6}$ ($\epsilon/\sigma^{3}/\tau$) for computational expedience. 

\begin{figure*}
\includegraphics[width=1.8\columnwidth]{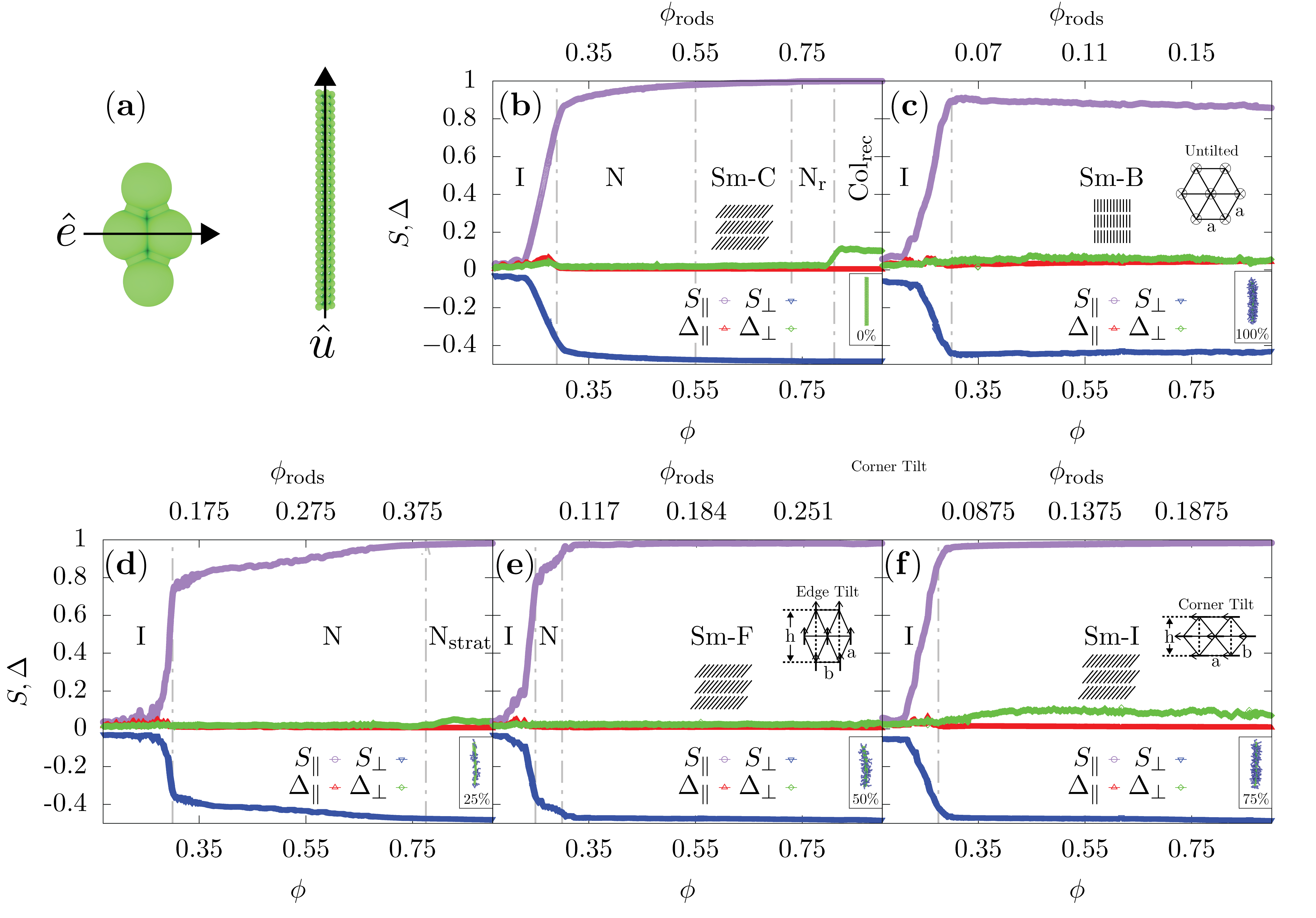}
\caption{\label{fig:order} (a) Cartoon schematics of the long and short-axis used in the definition of the global orientational order parameters. (b-f) Global orientational order parameters, using the long ($S_{\parallel}$, $\Delta_{\parallel}$) and short axes ($S_{\perp}$, $\Delta_{\perp}$) of the rods for all systems. Panels (b) and (c) represent naked ($\rho_{\mathrm{g}}=0$) and fully grafted ($\rho_{\mathrm{g}}=1$) rods respectively. Panels (d-f) represent intermediary grafting densities $\rho_{\mathrm{g}}=0.25,0.5$ and 0.75. The volume fraction of rods is denoted by $\phi_{\mathrm{rods}}$, $\phi$ represents the total volume fraction occupied by all components including both rods and grafted polymers. A single grafted rod is shown in the bottom right of each panel to illustrate the changing grafting density between systems.  
}
\end{figure*}

Before examining the phase-behaviour of each system, a series of order parameters must be introduced to examine the extent of LC ordering. To monitor orientational ordering of the long CNC axis the global tensor order parameter is defined as follows
\begin{equation}
 \boldsymbol{Q}_{\parallel}  = \frac{1}{N_{\mathrm{c}}} \sum_{i}^{N_{\mathrm{c}}}  \frac{3}{2} \langle \boldsymbol{\hat{u}}_{i}\boldsymbol{\hat{u}}_{i} \rangle  - \frac{1}{2}\boldsymbol{I} 
\end{equation}
where $\boldsymbol{Q}_{\parallel}$ is a traceless symmetric 2nd-rank tensor, the angular brackets a canonical average  
and $\boldsymbol{\hat{u}}_{i}$ is the unit vector spanning the end-to-end distance of the $i$-th rod.  In order to probe orientational order of the non-circular cross sections a similar tensor is defined and applied to the easy axis $\boldsymbol{\hat{e}}_{i}$ defined in  Fig. \ref{fig:order} (a) 
\begin{equation}
 \boldsymbol{Q}_{\perp}  = \frac{1}{N_{\mathrm{c}}} \sum_{i}^{N_{\mathrm{c}}}  \frac{3}{2} \langle \boldsymbol{\hat{e}}_{i}\boldsymbol{\hat{e}}_{i} \rangle - \frac{1}{2}\boldsymbol{I}  
\end{equation}
For both tensors the principal nematic order parameter $S$ is identified with the largest  eigenvalue $\lambda_{1}$ and the nematic director with its corresponding eigenvector \cite{de1993physics}. In order to assess the emergence of biaxial orientational order, a biaxial order parameter $\Delta = (2/3)(\lambda_{2} - \lambda_{3})$ is monitored in terms of the difference between the two minor eigenvalues $\lambda_{2,3}$ of each tensor.

In addition to the global orientational order parameters, the positional (or Smectic) ordering of the centre of mass (COM) of the rods is also examined. The Smectic order parameter $\Lambda$, is defined as follows, 
\begin{equation}
\Lambda =  \frac{1}{N_{\mathrm{c}}} \sum_{i}^{N_{\mathrm{c}}} \left \langle \cos{\bigg(\frac{2\pi m_{\parallel,i}}{l_{\mathrm{s}}}}\Bigg) \right \rangle
\end{equation}
where the sum is performed over the center-of-mass  (COM) coordinates of all $N_{\mathrm{c}}$ rods, $l_{\mathrm{s}}$ denotes the Smectic layer spacing and $m_{\parallel,i} = \boldsymbol{r}_{i}\cdot\boldsymbol{\hat{m}}$ represents the projection of the COM coordinates $\boldsymbol{r}_{i}$ of the $i-$th rod onto the Smectic layer normal $\boldsymbol{\hat{m}}$. Whilst this analysis is standard in Sm-A phases which are well aligned with the simulation cell i.e along one of the principal axes, tilted phases require knowledge of the layer normal. Thus we perform a sampling of vectors around the Nematic director $\boldsymbol{\hat{n}}$ within a conical space where sample vectors do not exceed a maximum angle of $30^\circ$ with $\boldsymbol{\hat{n}}$, this is combined with a sampling of the layer spacing $\delta l_{\mathrm{s}}\pm 6\sigma$ and the highest response taken as the Smectic order parameter. This allows the Smectic ordering to be examined in tilted phases without picking up other harmonics, for instance those related to in-plane ordering which will be discussed later. 

\section{Results and Discussion}

 In the first half of this section, the observed phase sequence and the phase diagram are mapped out for all systems in order of increasing grafting density using order parameters and system snapshots. The latter half is dedicated to characterising the detailed structure of each phase from the  static structure factor and other metrics including the Smectic tilt angle and layer spacing. At the very end, the layer hopping dynamics of rods within the Smectic phases is touched upon where the van Hove correlation function provides a clue as to what extent the colloids are confined to their layers. It also serves as a dynamical criterion to discriminate between Smectic phases with long-range in-plane periodicity and true (3d) crystalline order.   

\subsection{Phase Behaviour}
\label{sec:phasemapping}

The global Nematic order parameters for each system are shown in Fig. \ref{fig:order}, where both the short and long axes order parameter of the rods, $S_{\bot}$ and $S_{\parallel}$ respectively, have been monitored during each compression run, see the cartoon in Fig. \ref{fig:order} (a). In all cases, $\phi$ and $\phi_{\mathrm{rods}}$ denote the total volume fractions of all beads and volume fractions of only the beads comprising the backbone rods respectively. Beginning with the ungrafted system ($\rho_{\mathrm{g}}=0$), it is expected rods with a circular cross-section should follow the traditional Isotropic\textrightarrow Nematic\textrightarrow Smectic-A (I\textrightarrow N\textrightarrow Sm-A) phase-sequence that is commonly observed in steeply repulsive cylinder-shaped particles with sufficient anisotropy \cite{frenkel1988thermodynamic,bolhuis1997tracing,mcgrother1996re}. However the anisotropic cross-section of the rods considered here appears to induce much richer behaviour.

\begin{figure}
\includegraphics[width=\columnwidth]{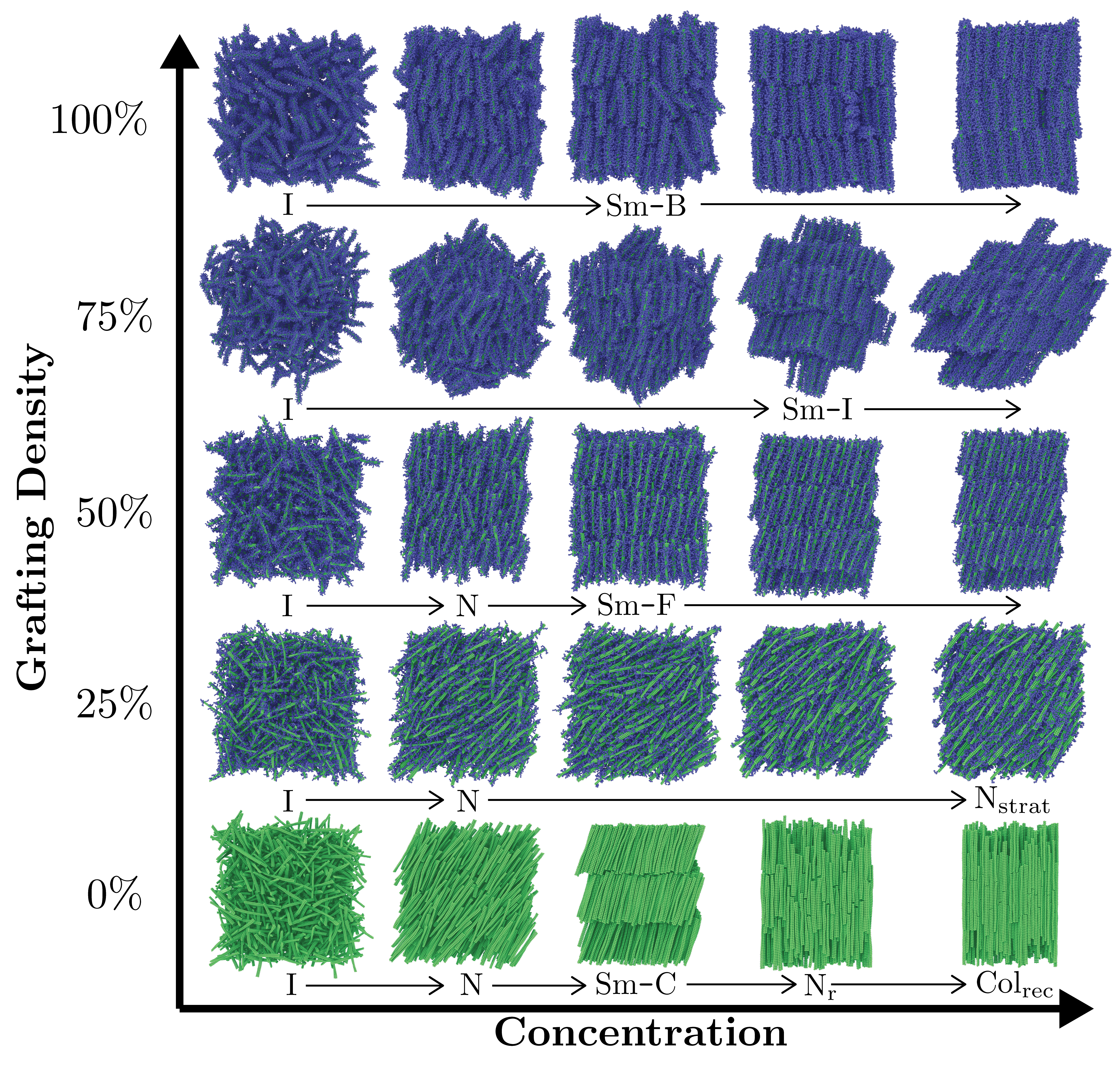}
\caption{\label{fig:snaps} System snapshots as a function of increasing concentration showing the different LC phase sequences. For each of the different systems corresponding to $\rho_{\mathrm{g}}=0,0.25,0.5,0.75$ and 1, the corresponding phase sequences are I\textrightarrow N \textrightarrow Sm-C \textrightarrow N\textsubscript{r} \textrightarrow Col\textsubscript{rec}, I\textrightarrow N, I\textrightarrow N\textrightarrow Sm-F, I\textrightarrow Sm-I and I\textrightarrow Sm-B respectively.}
\end{figure}

\begin{figure}
\includegraphics[width=\columnwidth]{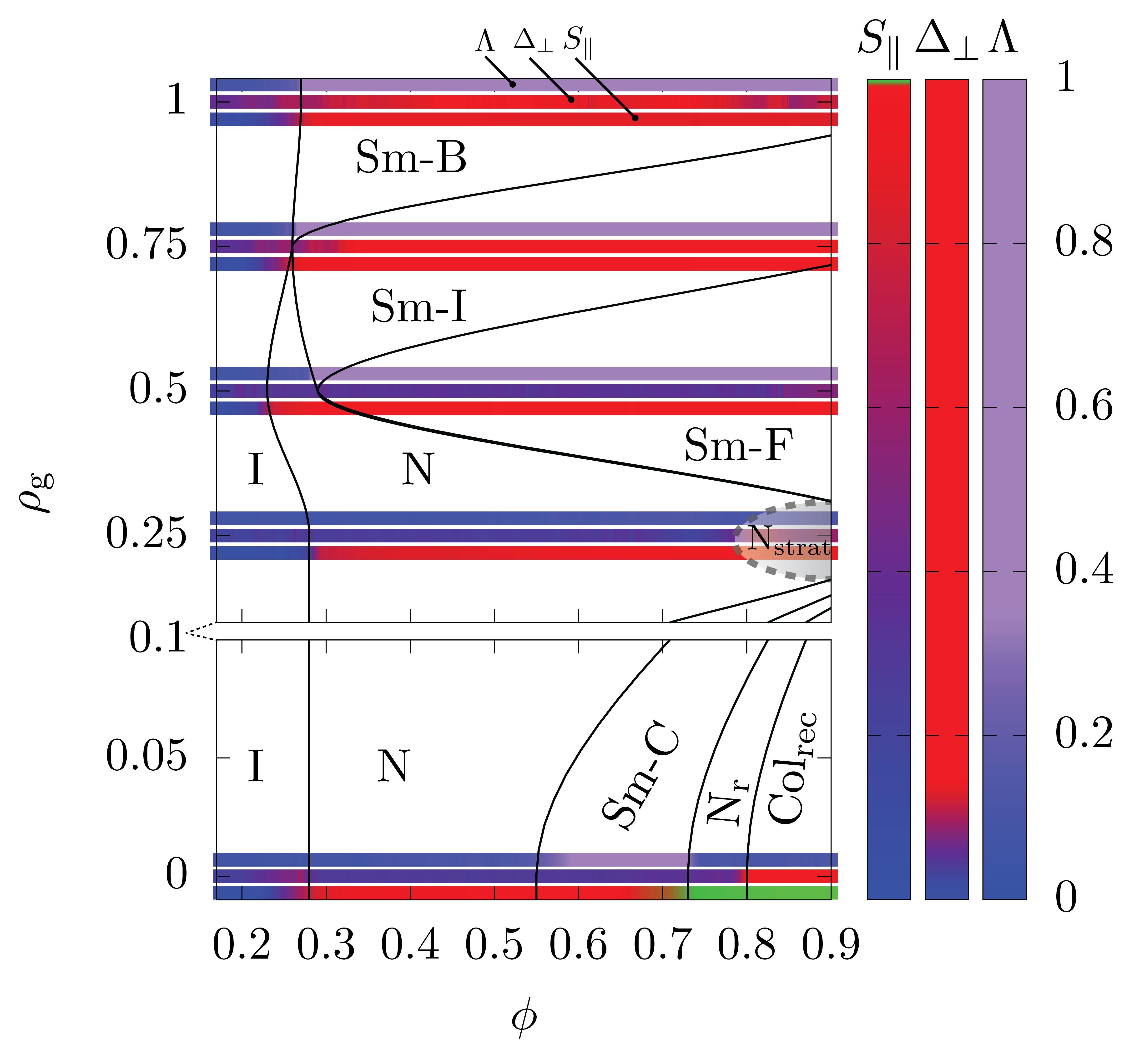}
\caption{\label{fig:phasediag} Phase diagram as a function of grafting density $\rho_{\mathrm{g}}$ and total occupied volume fraction $\phi$ comprising all colloid and polymer beads. The region between $\rho_{\mathrm{g}}=0$ and $\rho_{\mathrm{g}}=0.1$ is shown on a different scale to better reveal the small pockets at high concentrations, both panels are continuous. The bottom, middle and top colour bars correspond to the long-axis Nematic, short-axis Nematic and Smectic order parameters ($S_{\parallel}$, $\Delta_{\bot}$ and $\Lambda$) respectively. Lines are drawn through the respective jumps in the order parameters to approximate the phase boundaries. Note the color bars are scaled differently in order to highlight the jumping order parameters between different phases. 
}
\end{figure}

In Fig. \ref{fig:order} (b) a series of jumps are observed both in the long and short-axis order parameters at different concentrations. Initially the orientation of the rods in space is isotropic with small non-zero values in both order parameters until relatively high concentration where a sharp increase in both is observed around $\phi \approx0.28$. The critical packing fraction is in line with Onsager's \cite{onsager1949effects}  prediction  $\phi_{IN} \sim 4(D/L) \approx 0.27 $ for rigid hard rods taking an aspect ratio $L/D \approx 15$ and also agrees with values reported from Monte Carlo simulations for hard \cite{bolhuis1997numerical} and soft spherocylinders \cite{vink2005interfacial}.  In the thermodynamic limit $N_{c} \rightarrow \infty$, the transition from I\textrightarrow N is known to be a first-order one and should be accompanied by a sudden jump in the orientational order parameter. Evidence for the order of the phase transitions is provided by the hysteresis in the Nematic order parameter that we observe from decompression runs, see Figure S1 in the Supplementary Information. 
  The long-axis order appears globally aligned as indicated by $S_{\parallel}$ whereas the short-axis one is anti-nematic as reflected by a negative value $S_{\perp} <0$. Despite the weakly biaxial shape of the CNCs,  the biaxial order parameters associated with the long and short axis, respectively denoted by $\Delta_{\parallel}$ and $\Delta_{\perp}$, remain close to zero at low to moderate density which confirms that the global orientational symmetry of the LCs is uniaxial and not biaxial \cite{luckhurst2004missing,tschierske2010biaxial}. Upon further increasing the concentration, the  Nematic phase transitions continuously into a tilted Smectic-C (Sm-C) phase as shown by the snapshot in Fig. \ref{fig:snaps}. This is more clearly seen in Fig. \ref{fig:phasediag}, where the Smectic order parameter $\Lambda$ is overlaid with the long and short-axis order parameters to approximately map the phase boundaries as a function of concentration and grafting density. The Smectic order jumps at $\phi\approx0.55$, indicating substantial positional ordering. Since the CNC interactions are purely repulsive the observed tilt is likely stabilized by subtle excluded-volume effects imparted by the non-circular CNC cross-section \cite{somoza1988Nematic} and further compounded by weak twisting.  Comparing our scenario with the emergence of a Smectic A phase in systems of rigid hard (sphero)cylinders we find  that the Sm-C phase appears at a slightly larger packing fraction than $\phi \approx 0.45- 0.55 $ reported for cylinders \cite{bolhuis1997tracing,lopes2021phase}. The discrepancy could be due to the backbone fluctuations and non-circular cross section of our CNCs which cause their shape to deviate from mere rigid cylinders with a prescribed aspect ratio.

At even higher concentrations a small jump is observed in the long-axis order parameter as the Sm-C phase transitions into a reentrant Nematic (N\textsubscript{r})  phase at ($\phi\approx0.73$) in which rods appear stacked on top of one another with a complete loss of lamellar ordering seen in the Sm-C phase, see Fig. \ref{fig:snaps}. Since no in-plane positional ordering could be detected the global symmetry of this phase is Nematic. This transition is unusual as it signals an entropy-driven re-entrant melting at elevated rod concentrations whereas common knowledge has it that the order should increase successively upon approaching close packing. The mechanism behind the sudden loss of Smectic ordering remains unknown. It is notable that the transition is not accompanied by a jump in the short-axis order parameter, which could hint at small but abrupt changes in the backbone flexibility, driven by packing effects, playing a subtle role in the competition between different entropies. It is only at higher concentrations $\phi\approx0.8$ that a jump occurs in the short-axis when the rods begin to fully crystallize and tile the plane in what is later identified as a rectangular columnar phase (Col\textsubscript{rec}). A weak biaxial signal is picked up ($\Delta_{\perp} \sim 0.1$) which suggests an orientational symmetry breaking of the short axis across the plane perpendicular to $\boldsymbol{\hat{n}}$. 

In the next paragraphs, the effect of randomly grafting short oligomers with  length  $l_{\mathrm{p}}=4\sigma$ onto the surface of the rods and its influence on the phase-sequence is discussed. Beginning with a 25\% occupation of surface sites ($\rho_{\mathrm{g}}=0.25$), the long/short axis order parameters in Fig. \ref{fig:order} (d) are very different. When compared with the ungrafted rods, the phase-transition from I\textrightarrow N is both sharper and shifted to marginally higher overall concentrations. However, the critical rod concentration at the transition is about half the value found for naked rods. The Nematic order parameters, however, appear slightly reduced which can be explained by the patchy grafting of oligomers which changes the effective shape of the rods by rendering them less anisotropic. 
 
\begin{figure}
\includegraphics[width=\columnwidth]{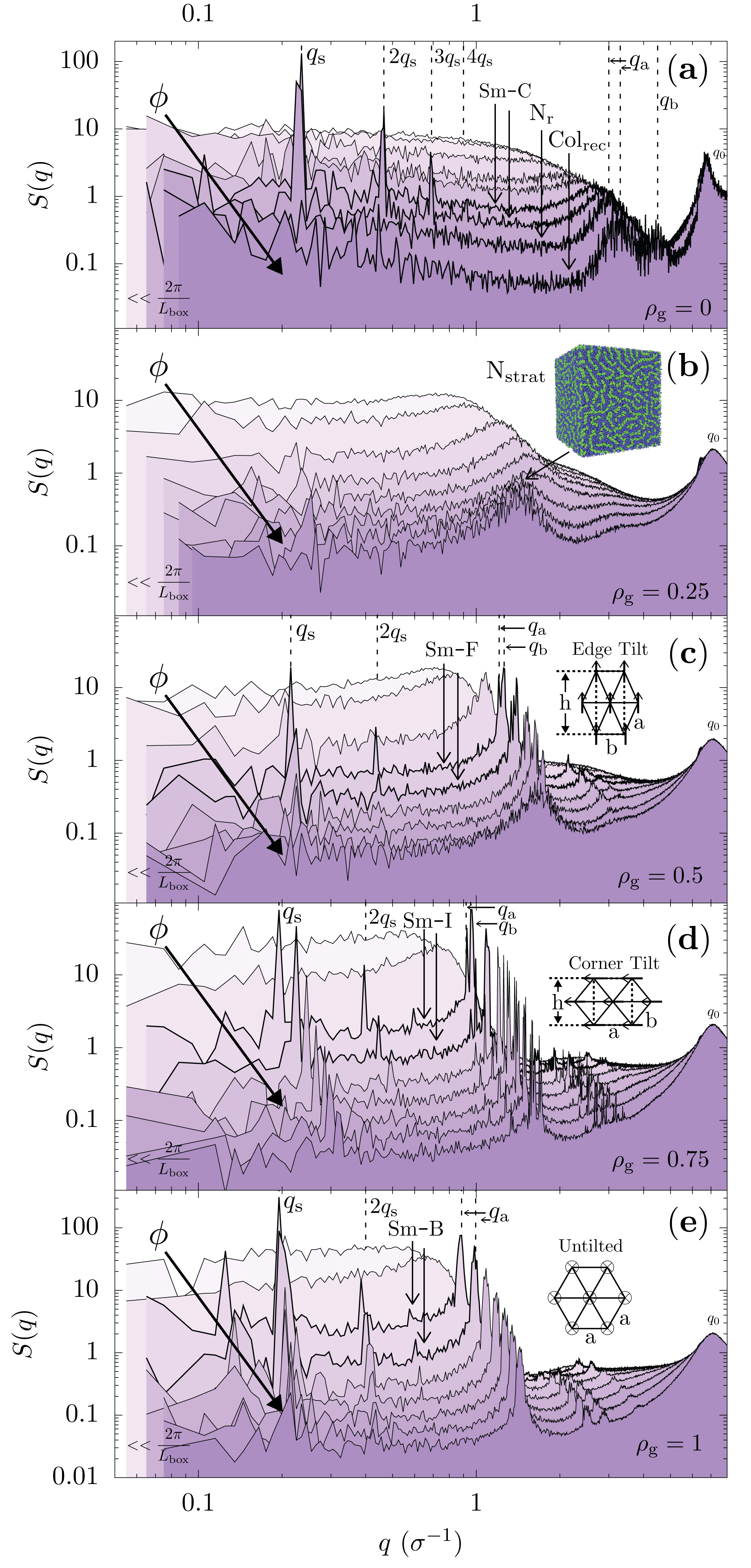}
\caption{\label{fig:strufa3d} Evolution of the static structure factor $S(q)$, with increasing concentration, panels (a-e) correspond to grafting densities  $\rho_{\mathrm{g}}=0,0.25,0.5,0.75$ and 1.0 respectively. The direction of increasing concentration $\phi$, is indicated by the inset arrow and each isobar is taken at intervals of $\delta\phi=0.1$ between $\phi=0.1$ and $\phi=0.9$. Bolder isobars are used to highlight complex Smectic phases, $q_{\mathrm{a}}$ and $q_{\mathrm{b}}$ correspond to the in-plane lattice parameters. The Smectic layer spacing $2\pi/l_{\mathrm{s}}$, is denoted by $q_{\mathrm{s}}$ and the lowest attainable $q$-value corresponds to $2\pi/L_{\mathrm{box}}$ as indicated in each panel. It is important to note that as the box changes shape, i.e. becomes more rectangular, some larger $q$ values can be obtained hence the non-monotonicity between isobars.
}
\end{figure}

This Nematic phase persists and no indication of Smectic ordering was observed even at very high concentration, see Supplementary Video 2. This is best reflected in Fig. \ref{fig:phasediag} where the Smectic order parameter picks up no signal that would point to long-range unidimensional positional order indicative of Smectic order. Sparse polymer grafting thus strongly suppresses positional ordering and stabilizes the Nematic fluid at the expense of Smectic order, at least within the range of pressures probed in our simulations. Similarly, no evidence of positional order developing perpendicular to the Nematic director is found, that would otherwise point to emergence of Columnar order.  
At higher concentrations ($\phi\approx0.6$), when the patchy rods are sufficiently dense, a `stratified' Nematic (N\textsubscript{strat}) appears as shown in Fig \ref{fig:strufa3d} where the backbones of the grafted rods appear to aggregate side-by-side to form long lamella trains separated by amorphous polymers. This can be seen clearly in the snapshots shown in Fig. \ref{fig:snaps} where the Nematic appears increasingly more globally aligned. This effect will shortly be further analyzed from the static structure factors. 

At a grafting density of 50 \% ($\rho_{\mathrm{g}}=0.5$) there is sufficient polymer surrounding the rod to recover the traditional phase-sequence. It is important to note as the rods are more heavily grafted, the effective aspect ratio of the rods is decreasing and it moves further towards a soft-rod description where the polymer coat begins to act as a uniform compressible medium. In this case the I\textrightarrow N transition is observed at much lower concentrations ($\phi\approx0.25$) and similarly disrupted with weaker orientational ordering than the ungrafted rods, see Figs. \ref{fig:order} (e) and \ref{fig:snaps}. With increasing concentration, the Nematic phase transitions gradually into a tilted Smectic-F phase (Sm-F) with in-plane rectangular order at relatively dilute conditions of around $\phi\approx0.3$, as discussed in detail later in the next Section. The conditions under which the Smectic phase forms are in stark contrast to the ungrafted rods which do not form a Smectic until around $\phi\approx0.55$ which agrees with what is expected for cylindrical mesogens of variable aspect ratio \cite{frenkel1988thermodynamic,bolhuis1997tracing,mcgrother1996re}. This suggests that grafting oligomeric polymers thus provides an effective means of stabilizing Smectic order at very low colloid content (about $10\%$).
Note that the tilt in the Sm-F phase is shallow but grows stronger with increasing concentration as illustrated in Fig. \ref{fig:snaps} (see Supplementary Video 3), the tilt angle is discussed in detail in the following Section.

With even higher grafting at 75 \% occupation ($\rho_{\mathrm{g}}=0.75$), the anisotropy of the rods decreases enough to fully destabilize Nematic order and induce a direct transition to a tilted Sm-I phase with in-plane rectangular order, see the following Section. Monte Carlo simulations for hard (sphero)cylinders \cite{bolhuis1997tracing,lopes2021phase}. report direct transitions from isotropic to Smectic A-type order if the aspect ratio drops below 3. Translating this to our case we may naively associate a lower bound effective aspect ratio $L/D \approx 1.5$  for densely grafted CNCs if the polymers are taken to be fully stretched and closely packed together.  As shown in Figs. \ref{fig:order} (e,f), the pronounced shoulder seen in both the short and long-axis order parameters in panel (e) has been replaced by a steep transition to Sm-I in panel (f). This is not unexpected for very short rods with low aspect ratios as is the case for heavily grafted rods. What is particularly striking in this case is that the tilt angle, most clearly seen in Fig \ref{fig:snaps}, appears extreme when compared to the system with only a 50\% occupation of surface sites, see Supplementary Video 4. This is discussed in detail in the upcoming section where the in-plane structure of all Smectic phases is characterised. The extreme director tilt appears to induce weak levels of biaxiality in the orientational order of the short axis as is noticeable in Fig. \ref{fig:order} (f).

The final system considers rods which are fully grafted with a 100\% occupation of surface sites ($\rho_{\mathrm{g}}=1$). Similarly to the previous system at 75\% grafting, the transition in Fig. \ref{fig:order} (c) proceeds directly to a complex Smectic phase but with key differences. Due to the complete uniform grafting of the rods, the rods become virtually identical in their dimensions and the system prefers a more uniform Sm-B phase with in-plane hexagonal order. This is seen most clearly in the snapshots in Fig. \ref{fig:snaps} where the system resembles a typical Smectic phase with the rod director aligned with the layer normal. Here it is important to note also a small grain-boundary defect is present in the system, not uncommon in simulations of Smectic forming rods \cite{milchev2019Smectic}, which partially disrupts the layering, see Supplementary Video 5. Within the range of parameters probed here, a conventional Sm-A phase is not observed, which would otherwise lack any long-range positional order perpendicular to the layer normal. This indicates that even heavily grafted rods still deviate considerably in their behavior from what is known from soft cylinder models \cite{earl2001computer, cuetos2002monte} or polymer-coated virus rods \cite{grelet2014hard,grelet2016soft} where the Sm-A phase continues to be the most prominent Smectic symmetry. From Fig. \ref{fig:phasediag} note that polymer grafting, provided dense enough, proves as an efficient means to stabilize Smectic order at low material content, given that the Smectic phase forms at $\phi \approx 0.3$ corresponding to a much lower CNC packing fraction of about $\phi_{\rm rods} \approx 0.1$. The transition towards Smectic order thus occurs at values that are way below the typical benchmark range  $\phi = 0.45 - 0.5 $  reported for hard (sphero)cylinders \cite{bolhuis1997tracing,lopes2021phase}. Another discrepancy between our model and rigid cylinders is that the onset of Nematic ordering seems rather insensitive to the degree of polymer grafting, whereas one would expect the critical packing fraction to systematically increase with $\rho_{\rm g}$ as the rods get fatter and their effective aspect ratio is reduced by the grafted polymers.

\subsection{Smectic Microstructure}\label{sec:characterise}

Characterizing the symmetry of the different Smectic phases requires knowledge of the static structure factor $S(q)$, which is defined as follows 
\begin{eqnarray}
S(q)=\frac{1}{N} \bigg \langle \sum_{i,j}^{N} \operatorname{e}^{\mathrm{i} \mathbf{q} \cdot (\mathbf{p}_{i} - \mathbf{p}_j)} \bigg\rangle_{|\mathbf{q}|=q\pm dq}
\label{e:strufa}
\end{eqnarray}
where the sum is performed over all $N$ beads in the system with coordinates $\mathbf{p}$ and a phase-angle average is performed by monitoring $q$-vectors of length $q\pm dq$.  A running average is applied in the interval $q_{p}=\pi/L_{\mathrm{box}}$ with bin size 0.01$\sigma^{-1}$. Only $q$-vectors compatible with the finite box size may be considered, hence the precision becomes increasingly poor as $q$ approaches the inverse box size \cite{meyer2002formation}. The static structure factor is shown in Fig. \ref{fig:strufa3d} for a series of different isobars between $\phi=0.1$ and 0.9. 

\begin{figure*}
\includegraphics[width= 1.5\columnwidth]{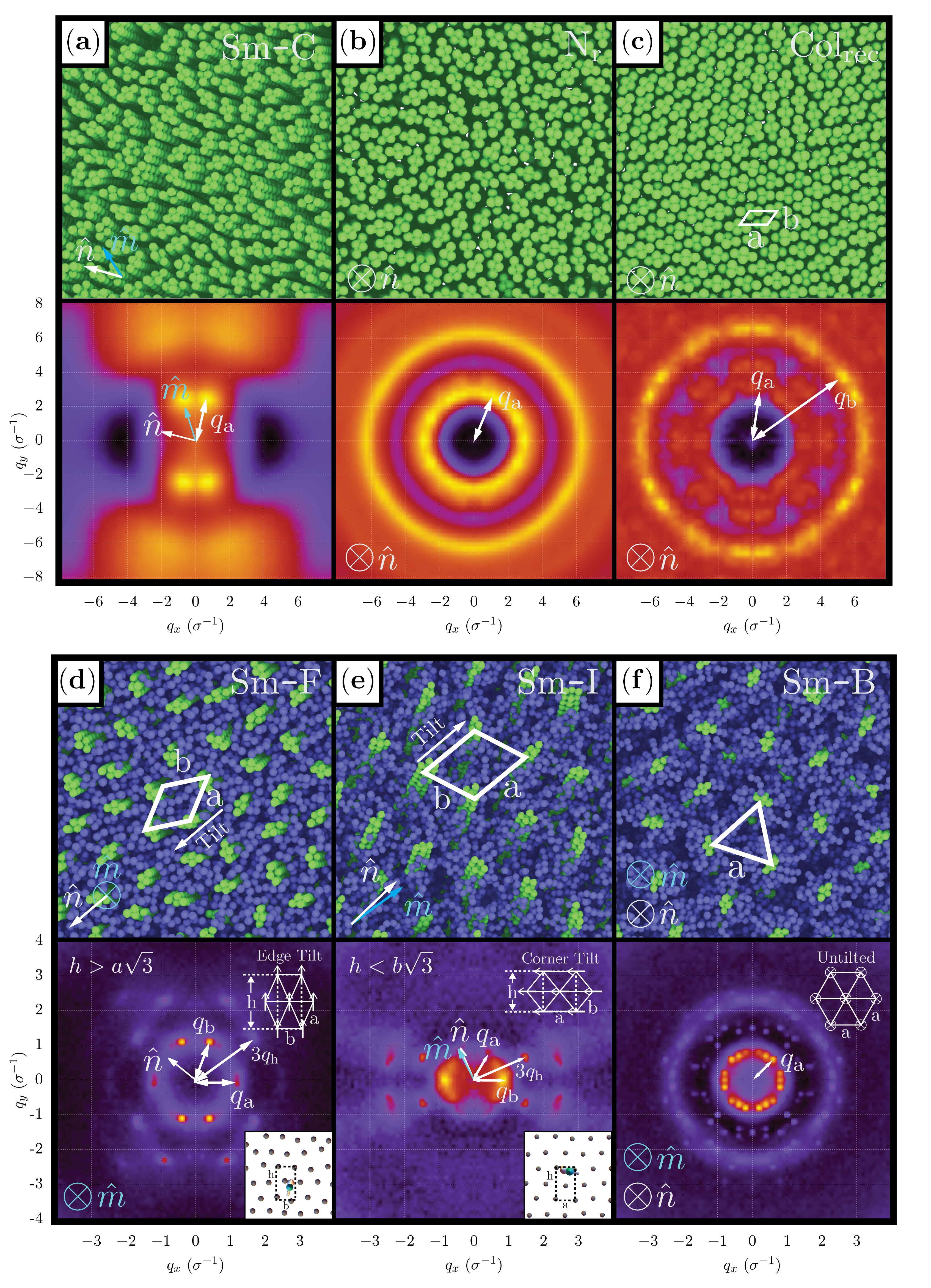}
\caption{\label{fig:strufa2d} Selected cut-throughs of the layer plane cross-sections perpendicular to $\hat{m}$  viewed along ${\hat{n}}$ of different phases and their accompanying 2d diffraction patterns. Panels (a-c) correspond to the SmC, N\textsubscript{r} and Col\textsubscript{rec} phases seen in the system of ungrafted rods, $\rho_{\mathrm{g}}=0$. Panels (d) and (e) correspond to the Sm-F and Sm-I phases in the $\rho_{\mathrm{g}}=0.5$ and 0.75 systems respectively. Note the diffraction patterns are slightly deformed due to the tilting with respect to the principal axes of the box. Panel (f) corresponds to the Sm-B phase in fully grafted rods $\rho_{\mathrm{g}}=1$ which is almost perfectly aligned with the box.}
\end{figure*}

The ungrafted rods show  a typical liquid like isobar devoid of peaks, which gradually transitions into a broader diffuse peak centred around $q=3\sigma^{-1}$ in the Nematic phase. Between $\phi=0.5$ and 0.6 a series of sharp peaks develop at 1,2,3 and 4 times the layer spacing $q_{\mathrm{s}}\approx0.2\sigma^{-1}$, corresponding to the Smectic layer spacing $l_{\mathrm{s}}\approx30\sigma$ which is consistent with the tilted layers and is on par with the rod length $l_{\mathrm{c}}=30\sigma$. Such sharp peaks accompanied by the presence of additional harmonics is indicative of long-range ordering within the simulation cell. In Fig. \ref{fig:strufa2d} (a), a snapshot of the cross-section of the box in real space (top) is shown along side the corresponding 2d diffraction pattern taken around one of the principal box axes in $q$-space (bottom). The pattern shows no sharp Bragg peaks indicating the absence of any long-range in-plane ordering. Instead the peaks are liquid-like and a splitting is observed with 4 bright diffuse peaks centred around $q=2.9\sigma^{-1}$. This splitting appears to be due to the tilting of the Smectic layer normal $\hat{m}$ and Nematic director $\hat{n}$ with respect to the principal box axis around which the in-plane diffraction pattern is formed. The diffuse spots occur in the $q$-range compatible with the diffuse peak at around $q=2.9\sigma^{-1}$ in Fig. \ref{fig:strufa3d} (a).

At higher concentrations, the Sm-C phase transforms into a reentrant Nematic phase N\textsubscript{r} which we identify due to the absence of long-range in-plane order in Fig \ref{fig:strufa2d} (b) where no sharp Bragg spots can be seen. Instead a series of diffuse rings are observed, in this case the principal box axis and Nematic director are aligned with one another resulting in the classic diffraction pattern of an isotropic liquid in the plane. The diffuse ring centred at $q\approx3\sigma^{-1}$ is consistent with the diffuse peaks occurring in Fig. \ref{fig:strufa3d} (a).  On further increasing the concentration, the reentrant Nematic phase transforms into a columnar phase with rectangular in-plane order as shown in Fig \ref{fig:strufa2d} (c), where multiple domain boundaries and differently oriented clusters can be seen, note the unit cell drawn in white. This explains the diffraction pattern which resembles that of a squashed hexagonal (rectangular) phase, with the additional peaks occurring due to multiple domains with different orientations as shown in the real space snapshot. In a perfect rectangular phase which is homogeneously aligned in-plane, 6 sharp peaks with two clear lattice parameters would be observed. These two distinct lattice parameters coincide with the central positions of the two diffuse peaks, $q_{\mathrm{a}}$ and $q_{\mathrm{b}}$ in Fig \ref{fig:strufa3d}(a). 
This multi-domain structure may result from the rectangular symmetry of the simulation box frustrating the unit cell, which could more easily tile a triclinic box as opposed to a rectangular one. It should be noted that the \Will{concentration} associated with the \Will{emergence of the }columnar phase is rather high and it may be possible that the discretized nature of our beaded CNCs (see Fig. \ref{fig:model}) could play a role in determining the in-plane structure at conditions where the rods are in close lateral proximity. At this point it is interesting to draw an analogy with condensed phases of filamentous virus rods where a similar loss of Smectic order at \Will{high concentrations} has been reported, except the favored symmetry is clearly hexagonal columnar rather than Nematic \cite{grelet2014hard}. In both cases, the corrugated effective colloid shape and the presence of backbone flexibility are believed to be subtle contributing factors that set these systems apart from simple rigid hard cylinders for which columnar order appears metastable \cite{lopes2021phase}. 


With the addition of grafting at $\rho_{\mathrm{g}}=0.25$ in Fig. \ref{fig:strufa3d} (b) the Isotropic phase gradually transitions into a Nematic as indicated by the pronounced diffuse peak around $q=1.5\sigma^{-1}$. The absence of sharp Bragg peaks indicates no long-range positional ordering is present. It is likely that the short number of side chains in this case suppresses crystallisation of the backbone rods preventing the formation of a complex Smectic. At high concentrations ($\phi\gtrsim 0.7$) the diffuse peak begins to sharpen suggesting pronounced short-ranged order is present within the Nematic. This is due to the what we call `stratification'  whereby local 1d lamella liquid trains form, see the inset snapshot in Fig. \ref{fig:strufa3d} (b). In this case the side chains group together exposing the naked face of the CNCs, which attract one another due to depletion. Speculatively, whilst the backbone rods are monodisperse in length, a source of polydispersity comes from the random grafting of chains onto the rods and this may prevent the formation of a globally aligned lamella (Smectic) phase, where the rods would otherwise lie perpendicular to the layer normal. It may be that different phases could be introduced through regular grafting of the different faces of the rods. 


At $\rho_{\mathrm{g}}=0.5$ the phase behaviour becomes very rich and the structure factor in Fig. \ref{fig:strufa3d} (c) reveals a complex structure. After the I\textrightarrow N phase transition, sharp Bragg peaks are observed both at low and high $q$ values. The peaks at $q_{\mathrm{s}}\approx0.2\sigma^{-1}$ correspond the Smectic layer spacing and the harmonics at 2 and 3 times  $q_{\mathrm{s}}$ indicate long-range ordering between successive layers. In addition, peaks at low $q$ result from the in-plane ordering, where two lattice parameters $q_{\mathrm{a}}$ and $q_{\mathrm{b}}$ can be observed and the presence of harmonics at higher $q$-values indicates the ordering is long-ranged within the correlation lengths achievable in this study. This is more clearly seen in the diffraction patterns in Fig. \ref{fig:strufa2d} (d) where the unit cell is drawn in for clarity. The layer normal nor the nematic director are fully aligned with the principal box axes which results in a partially deformed diffraction pattern, however the two lattice parameters and rectangular symmetry are still observed. Of the two spots the weaker reflection corresponds to the long-axis at $q_{\mathrm{a}}\approx1.21\sigma^{-1}$ and the brightest at $q_{\mathrm{b}}\approx1.26\sigma^{-1}$, note the Nematic director drawn in white points to neither of the sharpest Bragg peaks associated with $q_{\mathrm{a}}$ or $q_{\mathrm{b}}$. Instead the director points towards higher harmonics of $q_{\mathrm{h}}$ at 3$q_{\mathrm{h}}\approx2.0\sigma^{-1}$.  By definition tilting away from the principle axes of the unit cell is characteristic of the Sm-F phase \cite{benattar1983two,hanna2014charge} as denoted by the cartoon in Fig. \ref{fig:strufa3d} (d) where $h>a\sqrt{3}$.

\begin{figure*}
\includegraphics[width=1.8\columnwidth]{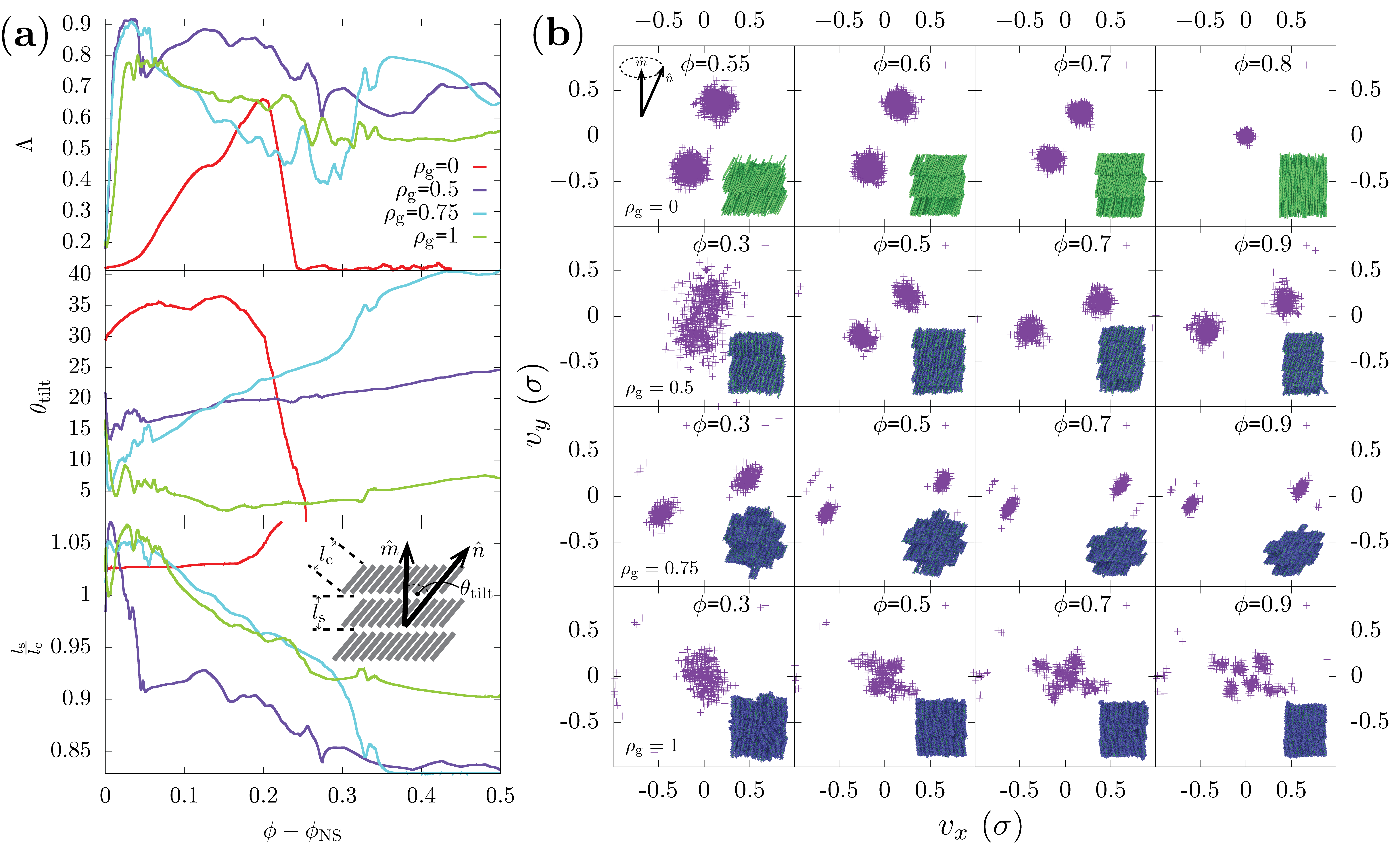}
\caption{\label{fig:tilt} (a) Evolution of the Smectic order parameter $\Lambda$, tilt angle  $\theta_{\mathrm{tilt}}$ and normalized layer spacing $l_{\mathrm{s}}/l_{\mathrm{c}}$ as a function of $\phi-\phi_{\mathrm{NS}}$ where $\phi_{\mathrm{NS}}$ is approximately the Nematic-Smectic transition concentration. Note $l_{\mathrm{c}}=30\sigma$ is the length of an ungrafted rod, see the cartoon inset into the bottom panel.  (b) Scatter plots showing the distribution of the end-to-end vectors of the rods $\boldsymbol{\hat{u}}$  about the layer normal $\hat{m}$ where $v_{x}$ and $v_{y}$ denote the resulting x and y components of the rods unit vectors after projection onto the plane perpendicular to the layer normal. Progressive snapshots show the evolution of tilt angle $\theta_{\mathrm{tilt}}$ and layer spacing $l_{\mathrm{s}}$ for $\rho_{\mathrm{g}}=$ 0, 0.5, 0.75 and 1 from top to bottom respectively.  
}
\end{figure*} 

The lattice parameters in this instance are very similar which makes it near impossible to identify in the real space snapshots due to thermal fluctuations and the tilting w.r.t. the axes of the box. This is shown clearly in the inset real space COM coordinates of a single Smectic layer (bottom right) where the globally averaged layer normal and Nematic director are drawn in blue and white respectively. As the tilting becomes more pronounced at higher concentration, the additional harmonics of the layer spacing disappear first due to surface roughening but the harmonics of the in-plane ordering at high-$q$ persist. Interestingly at very high concentrations some `stratification' appears analogous to the system in Fig. \ref{fig:strufa3d} (b) again due to depletion where the in-plane harmonics disappear and a single diffuse peak is observed centred around $q\approx1.6\sigma^{-1}$. It is worth pointing out that within the box sizes achievable here $\sim 80\sigma$, the correlation length achievable is very short. In view of the limited range of system sizes addressable within this study, the issue of whether the ordered Smectics display true long-ranged order or quasi-long-ranged  (hexatic) order in the thermodynamic limit cannot be resolved \cite{de1993physics,brock1989hexatic}. Given the complexity of our model system such a study would clearly be very challenging even with modern computational resources. 

Increasing grafting density further still to 75\% ($\rho_{\mathrm{g}}=0.75$) similar behaviour is observed as in the case of Fig \ref{fig:strufa3d} (c) but with key differences. Most notably the Nematic phase does not appear and the liquid like isobars proceed directly into a complex Smectic which is structurally different to the Sm-F phase. The peak corresponding to the layer spacing shifts to lower $q$, with an increased layer spacing and additional harmonics at low-$q$ indicate long-range order between successive Smectic planes. Peaks at high $q$-values indicate long-range order in-plane and two clear lattice parameters at $q_{\mathrm{a}}=0.92\sigma^{-1}$ and  $q_{\mathrm{b}}=0.95\sigma^{-1}$ are observed. The tilting is particularly strong in this case and results in a substantial deformation of the diffraction pattern seen in Fig. \ref{fig:strufa3d} (d). Here the brightest spot corresponds to $q_{\mathrm{b}}$ and the weaker spot to $q_{\mathrm{a}}$ indicating a short-axis tilt indicative of Sm-I, where the director points conclusively towards one of the sharp Bragg peaks. This is further confirmed when examining the position of the higher harmonics of $q_{\mathrm{h}}$ at 3$q_{\mathrm{h}}\approx1.54\sigma^{-1}$ indicating $h<b\sqrt{3}$ and thus Sm-I symmetry. With increasing concentration these peaks move apart as the in-plane packing becomes more rectangular and the harmonics of the layer spacing become less pronounced consistent with extreme tilting.  Contrary to the previous case, the Sm-I phase persists even at high concentration with no stratification being observed due to sufficiently dense grafting.

With 100\% grafting ($\rho_{\mathrm{g}}=1$), Fig. \ref{fig:strufa3d} (e) shows no apparent isobars characteristic of a Nematic and instead proceeds directly to a Smectic. The layer spacing at $q_{\mathrm{s}}$ moves to marginally lower $q$ compared to the Sm-F and Sm-I phases in Fig \ref{fig:strufa3d} (c) and (d) with additional harmonics present indicating long-range order. The peaks at high-$q$ are substantially different with one single peak at $q_{\mathrm{a}}\approx0.9\sigma^{-1}$ indicating only a single lattice parameter is present characteristic of a Sm-B with additional in-plane harmonics. This is no more apparent than in Fig. \ref{fig:strufa2d} (f) where the diffraction pattern shows bright Bragg spots in a near perfect hexagon. In this case the Nematic director and layer normal are almost perfectly aligned with the principal box axes. Interestingly instead of 6 bright spots, 18 are present at different angles, this is due to the presence of smaller Sm-B domains inside the box with different orientations as previously highlighted, which is both reflected in the snapshots in Fig. \ref{fig:snaps} and the order parameter profiles in Fig. \ref{fig:order} (c). At higher concentrations harmonics at high and low $q$ persist but reduce slightly due to the additional domains which partially deform the Sm-B layering inducing some small average global tilt w.r.t the simulation cell, see Supplementary Video 5.

In Figs. \ref{fig:tilt} (a) and (b) the tilt angle and layer spacing is examined more closely between systems. For ungrafted rods, the Sm-C phase forms tilted with $\theta_{\mathrm{tilt}}\sim30^\circ$ initially with a layer spacing $l_{\mathrm{s}}/l_{\mathrm{c}}\sim1.025$ which is marginally larger than the rod length. This is likely due to surface roughening between the layers of the Sm-C. At the peak of the Smectic order parameter $\Lambda$, $\theta_{\mathrm{tilt}}\sim25^\circ$ and the layer spacing remains almost constant. The surface roughness between layers appears to reduce prior to entering the reentrant Nematic phase as shown in Fig. \ref{fig:tilt} (a). Whilst the Smectic order continuously increases as the Sm-C phase forms, the grafted systems show a near discontinuous jump into complex Smectics. 

The behaviour of the grafted rods appears radically different between systems. In the Sm-F phase at 50\% grafting the phase forms tilted initially with a very small average tilt angle $\theta_{\mathrm{tilt}}\sim15^\circ$ and a layer spacing $l_{\mathrm{s}}/l_{\mathrm{c}}\sim1.05$, this is suggestive of an untilted Sm-A type structure  before a strong global tilting sets in. The correlation between layers appears weaker in the snapshot in Fig. \ref{fig:tilt} (a) initially and is consistent with the lack of harmonics of the layer spacing and hexagonal lattice in Fig. \ref{fig:strufa3d} around $\phi=0.3$ (3rd isobar) immediately preceding the Sm-I, suggesting only short-ranged ordering is present in-plane. In Fig \ref{fig:tilt} (b) the distribution of orientations of the rods about the layer normal is examined. In the panels corresponding to Sm-F at $\phi=0.3$ a diffuse spot is present suggesting local but no global tilt which somewhat resembles the diffuse-cone picture of de Vries-type Smectic phases \cite{de1979description}. It is possible that a small pocket of Sm-A could exist in the phase diagram surrounding this region. As the concentration is further increased, the tilt angle increases  almost linearly towards a maximum of $25^\circ$ accompanied by a decrease in layer spacing where $l_{\mathrm{s}}/l_{\mathrm{c}}\sim0.84$. The scatter plots showing the projected orientations $\boldsymbol{\hat{u}}$ of the individual rods in Fig \ref{fig:tilt} (b) reflect this with the two identical sharp spots in the upper right and lower left quadrant confirming apolar orientational order.  

In the Sm-I phase at 75\% grafting, the tilt angle appears to start lower than the Sm-F around $\theta_{\mathrm{tilt}}\sim5^\circ$ with a comparable layer spacing $l_{\mathrm{s}}/l_{\mathrm{c}}\sim1.05$. The tilt angle increases almost linearly with increasing concentration towards a maximum of $\theta_{\mathrm{tilt}}\sim 40^\circ$ which is the highest tilt angle observed in all of the systems reported here. Speculatively, the extreme tilt in this case likely arises due to strong depletion effects which decrease the in-plane spacing between the rods, requiring extreme tilting to free up additional free volume for the chains at the end of the rods to explore. This is clearly seen in Fig. \ref{fig:tilt} (b) where the sharp spots move further apart, which shows extreme tilting w.r.t the layer normal. The fully grafted system does not tilt and the small average global non-zero tilt $\theta_{\mathrm{tilt}}\sim5^\circ$ is due to the presence of multiple Sm-B domains within the box. This is clearly seen in Fig \ref{fig:tilt} (b) where multiple sharp spots appear indicating the presence of 3 separate SmB domains, i.e. 3 pairs of 2. Note back to Fig. \ref{fig:strufa2d} and the 18 sharp spots as opposed to 6 in the diffraction pattern in panel (f) suggesting 3 domains with different orientations, see Supplementary Video 5. 

\subsection{Layer Hopping Dynamics} 
\label{sec:hopping}

\begin{figure*}
\includegraphics[width=1.8 \columnwidth]{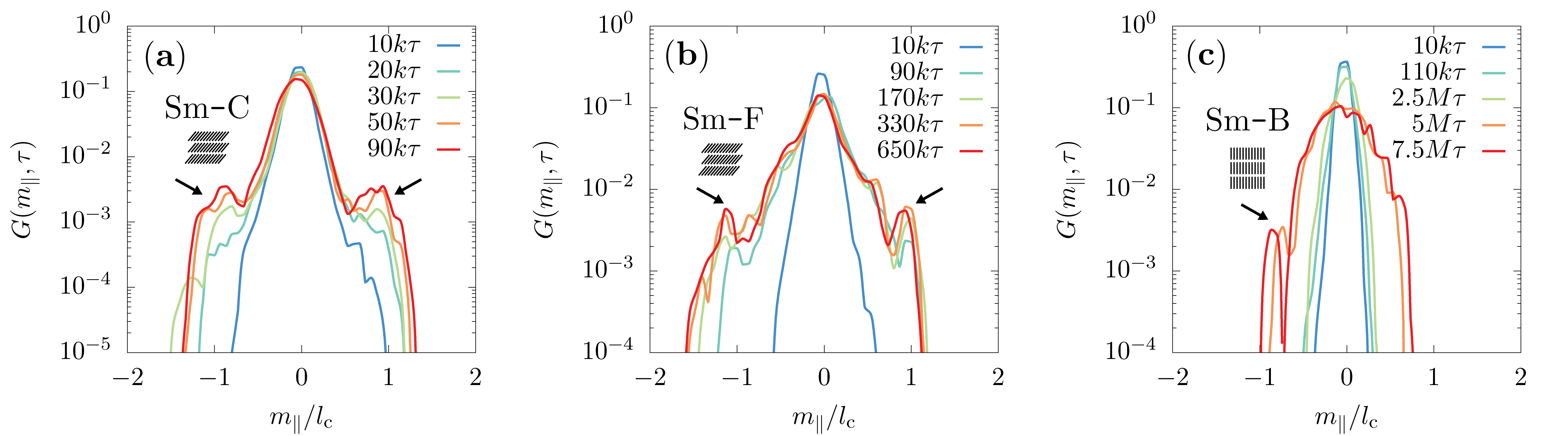}
\caption{\label{fig:vanhove} Van Hove correlation functions $G(m_{\parallel},\tau)$ for (a) ungrafted rods ($\rho_{\mathrm{g}}=0$) in the Sm-C phase, (b) partially grafted rods ($\rho_{\mathrm{g}}=0.5$) in the Sm-F phase and (c) fully grafted rods ($\rho_{\mathrm{g}}=1.0$) in the Sm-B phase. The displacement $m_{\parallel}$ along the layer normal is normalized by the rod length $l_{c}$. Curves are depicted at varying times during the compression run, the distinct shoulders highlighted by the inset arrows indicate diffusive barriers between the lamellae of the Smectic phases. 
}
\end{figure*}

To assess the mobility of the rods along the layer normal of the various Smectics, the self-part of the Van Hove correlation function is examined and is defined as \cite{lettinga2007self,ferreiro2018long}
\begin{equation}
    G(m_{\parallel},\tau)=\frac{1}{N_{\mathrm{c}}}\Bigg\langle\sum_{i=1}^{N_{\mathrm{c}}} \delta[m_{\parallel}+m_{\parallel,i}(0)-m_{\parallel,i}(\tau)]\Bigg\rangle_{\tau\pm\delta\tau}
\end{equation}
where $m_{\parallel,i}$ denotes the  position of the $i$-th rod projected along the layer normal $\boldsymbol{\hat{m}}$. The angular brackets indicate the time average over a short time window $\tau\pm\delta\tau$, where $\delta\tau=25\tau$.

Fig. \ref{fig:vanhove} shows the correlation functions for 3 sample phases, Sm-C, Sm-F and Sm-B. In panel (a), it is apparent that two shoulders are present (see inset arrows) indicative of the lamella structure of the Smectic phase and hopping of rods between layers. The time to the appearance of the first shoulder appears short $\sim30k\tau$. To establish a reference timescale, note that the diffusion coefficient of an ungrafted rod in the isotropic phase is $D=4.1\times10^{-3}$ $(\sigma^{2}/\tau)$, see Fig S3. Thus the time for a rod to travel its own length ($\sim 30 \sigma$) can be estimated as $\tau_{\rm D} = \sigma^{2}/D \approx 7k\tau$.
Thus the fastest hopping dynamics are approximately 3 times slower than the typical time of an ungrafted rod to diffuse its own length in the isotropic phase. In panel (b) grafted rods in the tilted Sm-F phase demonstrate similar behaviour with clear shoulders.  However the time to the first hopping events appears slower $\sim90k\tau$ which is not unexpected due to the larger number of beads per molecule. With fully grafted rods in the Sm-B phase in panel (c), the shoulders take a much longer time to appear $\sim 5M\tau$ but are nonetheless present.  This is not unexpected since fully grafted rods are 5$\times$ heavier than ungrafted ones. Furthermore, the appearance of shoulders provides evidence that the rods are not dynamically confined to a single layer but exhibit longitudinal mobility through hopping-type diffusion. This is a hallmark of Smectic liquid crystalline order and suggests that the systems reported here are truly Smectic phases rather than their 3d crystalline counterparts i.e. Sm-G or Sm-J crystals encountered in thermotropic liquid crystals \cite{coates2019thermotropic}. From a structural point of view, however, finite box-size effects prevent us from making any conclusive statements about the long-ranged ordered nature of the Smectic phases observed.


\section{Summary and Conclusions} 

Inspired by recent advances in polymer-functionalised cellulose nano-crystals, a simulation study of the effect of polymer grafting on the liquid crystalline (LC) self-assembly of stiff colloidal rods grafted with oligomeric polymers has been performed. The contour length of the polymers comprises only a fraction of the rod length so that the dressed rods retain a distinct anisotropic effective shape. Compared to naked rods, the effective interaction between the grafted colloids is strongly modified by the presence of the polymer and can be controlled by changing the surface density of the randomly grafted polymers. For simplicity, all interactions were purely repulsive so that all phase transformation were purely entropy-driven.

The coarse-grained (CG) colloid-polymer model developed here allowed access to the length scales required to grow LC phases in simulations of monodisperse CNCs with polydisperse grafting of freely-flexible polymer chains. Using a slow continuous-compression protocol, the spontaneous ordering of CNCs with different grafting densities could be observed and was discovered to be very rich forming a whole host of complex Smectics including Sm-C, Sm-F, Sm-I and Sm-B pointing to as yet undiscovered LC behaviour in grafted nanorods. Ungrafted rods also appear to exhibit unexpected phases including a reentrant Nematic and rectangular Columnar phase. Using different metrics including the long and short-axis Nematic and Smectic order parameters, the phase diagram could be approximately mapped as a function of concentration $\phi$ and grafting density $\rho_{\mathrm{g}}$ revealing that grafting density can drive the formation of different Smectic phases through depletion forces. Characterising their structure using 3d and 2d structure factors and real space snapshots revealed Sm-C, Sm-F, Sm-I and Sm-B phases can be induced by increasing grafting density. Interestingly, the formation of complex Smectics can be entirely suppressed at very low grafting densities. By monitoring the tilt angle and layer spacing it was found that de Vries like behaviour is generally not observed but a small pocket of de Vries like Sm-A could exist at intermediary grafting densities proceeding the Sm-I phase. By monitoring the dynamics of the rods, through the self-part of the Van Hove function, distinct hopping events were also observed between the lamella structure of the Smectic phases. This revealed an increased hopping time for more heavily grafted rods suggesting that the observed Smectic LC phases lack full 3d crystalline order.  

The CG model and protocol developed here now provides an opportunity to study the formation of new complex LC phases in polymer grafted nanorods and the effect of different metrics including polymer/rod length dispersity, chain length, grafting density and type i.e. targeted faces or regularly distributed chains and the influence of temperature or concentration on the phase diagram and even the dynamics of the rods in the host phases. Since CNCs are natively polydisperse in width and length, a natural extension of the present model would be to consider uniformly grafted rods with polydisperse lengths which may demonstrate pronounced Columnar as opposed to Smectic ordering given that the latter is known to be easily disrupted by length dispersity \cite{mederos2014hard}. In this regard it is hoped that this study will inspire further experimental work on size-purification of CNC suspensions, in an effort to realize CNC-based tilted Smectic materials which could find applications as photonic materials \cite{dkabrowski2004new,peddireddy2013lasing}. The impact of CNC backbone chirality, not considered in our study, could also be explored by supplementing our CNC model with a chiral symmetry breaking potential that favors one twisting direction over the other.  Finally, the resulting LC structures may be cross-linked to form gels and their suitability for use in devices such as biomimetic actuators \cite{ianiro2023computational} or LC displays tested. Work in these directions is currently underway and will be reported in a forthcoming publication. 


\section*{Acknowledgement}

This work was supported by the European Innovation Council (EIC) through the  {\em Pathfinder Open} grant ``INTEGRATE"  (no. 101046333).
 We gratefully acknowledge access to the HPC resources of IDRIS under the allocation 2022-A0130913823 made by GENCI.  The authors are grateful to  Patrick Davidson and Hendrik Meyer for helpful discussions.





\bibliography{fall_et_al.bib}

\newpage

\section*{Supplementary info}

\subsection*{Decompression Simulations}

\begin{figure}
\includegraphics[width=\columnwidth]{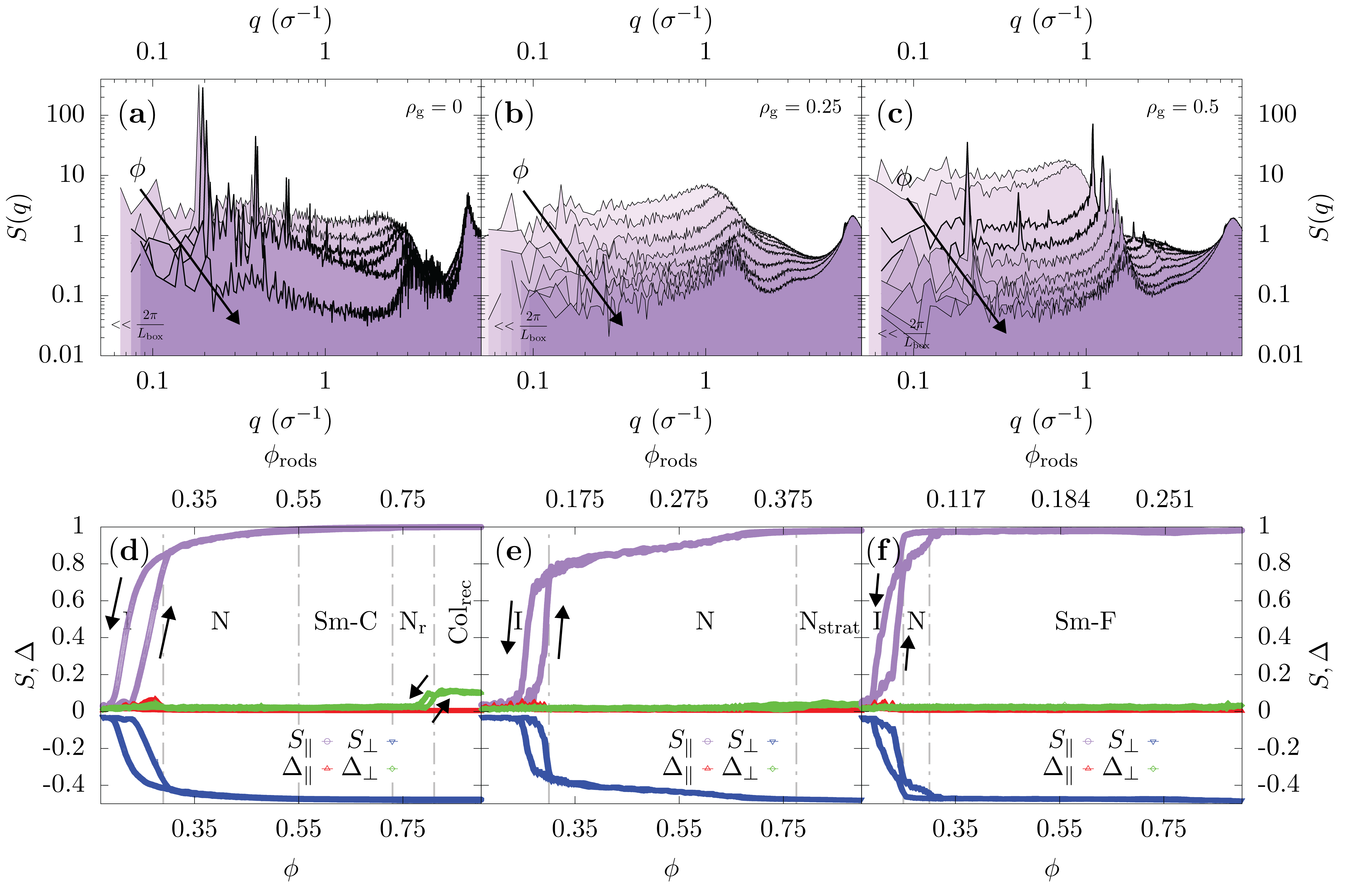}
\caption{Order parameters and structure factors during decompression runs for $\rho_{\mathrm{g}}=0,0.25$ and 0.5 systems. (a-c) Evolution of the static structure factor $S(q)$, with decreasing concentration. The direction of increasing concentration $\phi$, is indicated by the inset arrow and each isobar is taken at intervals of $\delta\phi=0.1$ from $\phi=0.9$ until the system reaches the isotropic phase. Bolder isobars are used to highlight complex Smectic phases. (d-f) Global orientational order parameters, using the long ($S_{\parallel}$, $\Delta_{\parallel}$) and short axes ($S_{\perp}$, $\Delta_{\perp}$) of the rods for all systems. The volume fraction of rods is denoted by $\phi_{\mathrm{rods}}$, $\phi$ represents the total volume fraction occupied by all components including both rods and grafted polymers as defined in the main manuscript. Up or down arrows indicate compression or decompression respectively.}
\label{fig:decomp}
\end{figure}

In order to assess the robustness of our simulations, additional decompression simulations were performed starting from high concentrations $\phi\sim0.9$ following the same procedure outlined in the main manuscript only in reverse. This was performed for systems with grafting density $\rho_{\mathrm{g}}=0,0.25$ and 0.5 to check the same phases were observed on decompression. In Fig. \ref{fig:decomp} the static structure factor $S(q)$ for each system is shown alongside the global Nematic order parameters on decompression. Both quantities provide a measure of the positional ordering and extent of hysteresis behaviour at the phase transitions between LC phases.

For $\rho_{\mathrm{g}}=0$, on decompression at the Col\textsubscript{rec}$\rightarrow$N\textsubscript{r} transition the structure factor in panel (a) shows the two diffuse peaks of the grainy Col\textsubscript{rec} phase disappearing and merging into a single diffuse peak at high-$q$ confirming the change in symmetry. Hysteresis behaviour is also observed in the biaxial order parameter (green curve) in panel (d) further demonstrating the reversibility of the transition. At the N\textsubscript{r}$\rightarrow$Sm-C transition, the Nematic order parameter strongly overlaps that of obtained during the compression run. The structure factor in panel (a) shows more intense peaks at low-$q$ indicating an improvement of the Sm-C order on decompression, due to rearrangement, with some additional harmonics at higher-$q$ values. The system gradually transitions into the N phase with considerable hysteresis being observed at N$\rightarrow$I, which is indicative of a first-order transition. Note the width of the hysteresis loop gives some indication of the size of the biphasic I$+$N gap.

At $\rho_{\mathrm{g}}=0.25$, on decompression the order parameters strongly overlap in panel (e) and the system de-`stratifies' on decompression as indicated in panel (b) by the gradual shift of the diffuse peak at intermediary $q\sim1.5$ to lower-$q$ values and it becomes gradually more diffuse as it does so, transitioning back to the traditional N phase.  We conclude that the stratification cross-over is not subject to hysteresis and happens at the same $\phi$ as on compression.  Considerable hysteresis being observed at the N$\rightarrow$I transition similarly to the naked rods. 

In the $\rho_{\mathrm{g}}=0.5$ system, on decompression, the order parameters strongly overlap in panel (f) until the Sm-F$\rightarrow$N transition and the peaks of the Sm-F phase are recovered in panel (c). Note the peaks are more intense due to some improvement of the Sm-F order on decompression and the removal of defects, i.e. rods between layers. At the Sm-F$\rightarrow$N transition considerable hysteresis is observed and the higher value of $S_{\parallel}$ indicates an improvement of the N ordering on decompression which is further confirmed by the more intense Nematic like shoulder in the isobar in panel (c) at high-$q$ after the Sm-F peaks disappear compared to the compression runs, see Fig 5(c) in the main manuscript. The system then transitions sharply to the I phase with considerable hysteresis being observed at the N$\rightarrow$I transition similarly to previous systems indicating the first-order nature of this transition.

\subsection*{Persistence Length Estimation}
In order to illustrate the varying persistence lengths of the ungrafted rods at different concentrations, the intra-rod orientational correlation functions is defined as follows
\begin{equation}
    P_{1}(s)=\langle\hat{u}_{i}\cdot\hat{u}_{j}\rangle_{\mid i-j\mid=s}
\end{equation}
where $\hat{u}_{i,j}$ correspond to two bond vectors along one of the 4 stems comprising the rods which are $s$ monomers apart, defining an effective contour. This correlation function decays approximately exponentially and may be used to estimate the persistence length $l_{\mathrm{p}}$ of the rods.
\begin{figure}
\includegraphics[width=\columnwidth]{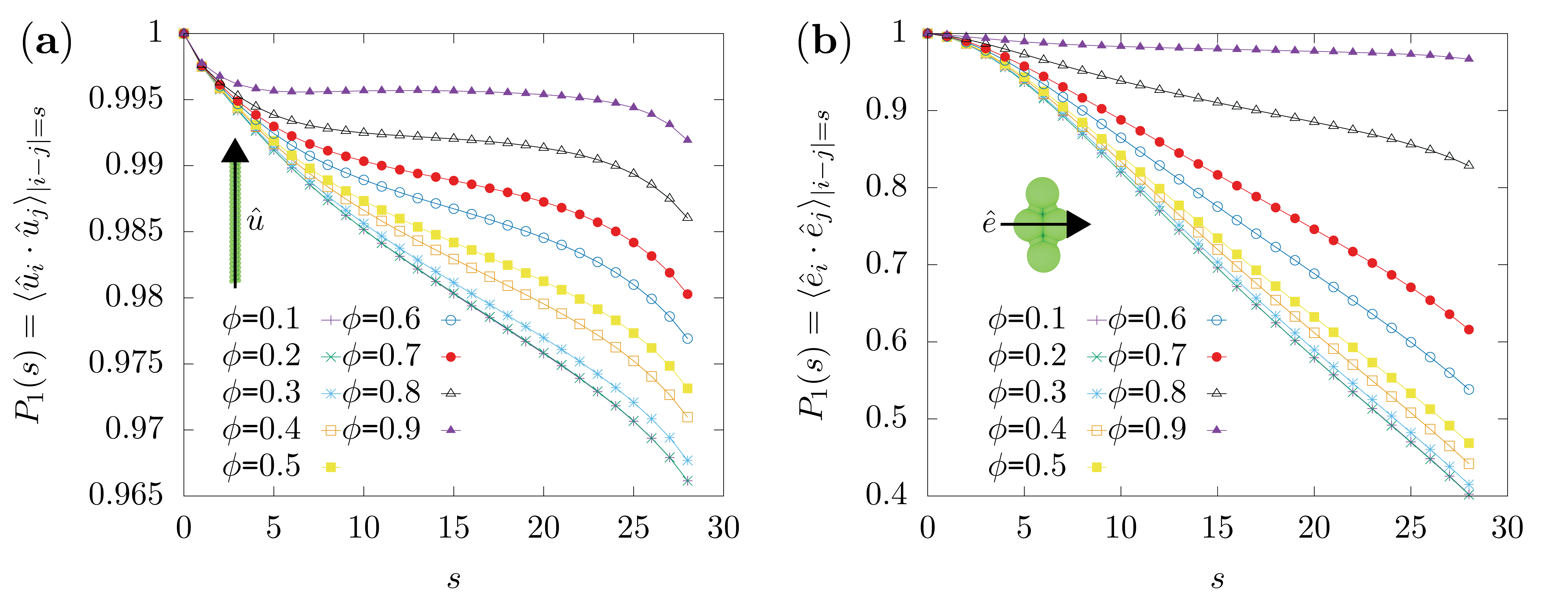}
\caption{Intra-rod orientational correlation function (OCF) $P_{1}(s)$ at different concentrations during the compression simulations. Panels (a) and (b) show the OCF of the long and short axis bead-to-bead bond vectors $\hat{u}_{i}$ and  $\hat{e}_{i}$ respectively. }
\label{fig:persist}
\end{figure} 
In Fig \ref{fig:persist} the OCF has been calculated for both the short and long axis fluctuations of the rods and reveals. \Rik{The plots clearly demonstrate that the rods are not perfectly rigid but exercise weak conformational changes. The correlation of the cross-section vectors [Fig. \ref{fig:persist}(b)] drops faster which means the rods exhibit considerable non-chiral twist fluctuations. In both cases, the decay of $P_{1}$ weakens when the packing fraction increases which indicates that the rods become stiffer in crowded conditions as expected.}

\subsection*{Mean Squared Displacement \& Diffusion Coefficient}

\begin{figure}
\includegraphics[width=0.8\columnwidth]{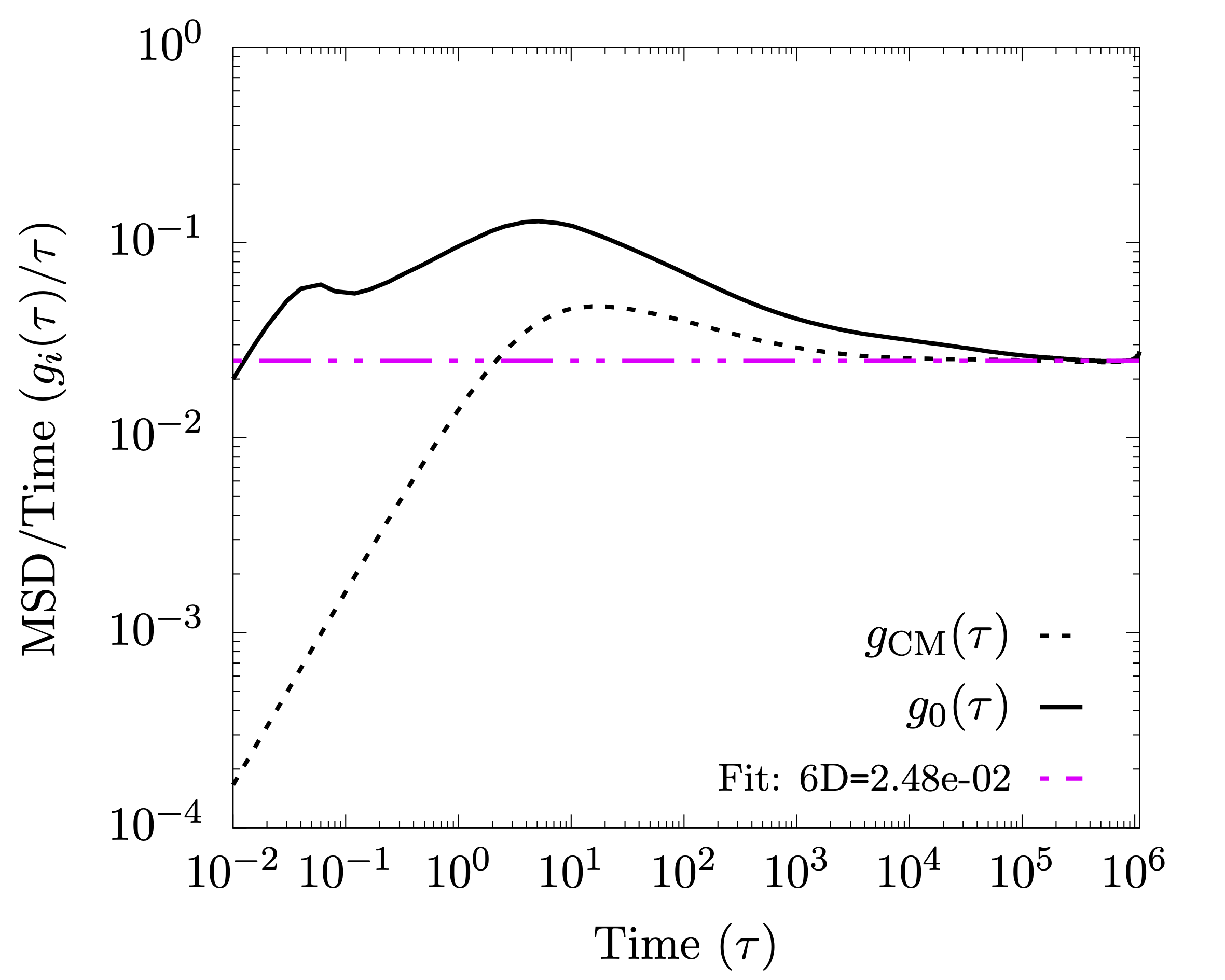}
\caption{Mean squared displacement of all rod monomers $g_{0}(t)$ and the centre-of-mass $g_{\mathrm{CM}}(t)$ of the rods as a a function of time. The pink dashed line indicates a linear fit to the diffusive regime at long times and is given in units of $\sigma^2/\tau$.  }
\label{fig:diffcoeff}
\end{figure} 

To assess the typical timescales of diffusion of the rods in the isotropic phase the mean squared displacement (MSD) of the beads comprising the rods is defined as follows

\begin{equation}
    g_{0}(t)=\frac{1}{N}\sum_{i}^{N}\langle\mid\boldsymbol{r}_{i}(t)-\boldsymbol{r}_{i}(0)\mid^{2}\rangle
\end{equation}
where $\boldsymbol{r}_{i}(t)$ is the position of the particle $i$ at time $t$ and the sum runs over all $N$ rod monomers. The MSD of the centre-of-mass (COM) of the rods is similarly defined as
\begin{equation}
    g_{\mathrm{CM}}(t)=\frac{1}{N_{\mathrm{c}}}\sum_{i}^{N_{\mathrm{c}}}\langle\mid \boldsymbol{r}_{\mathrm{CM}_{i}}(t)-\boldsymbol{r}_{\mathrm{CM}_{i}}(0)\mid^{2}\rangle
\end{equation}
where $\boldsymbol{r}_{\mathrm{CM}_{i}}(t)$ is the position of the COM of an arbitrary rod at time $t$ and the sum runs over all $N_{\mathrm{c}}$ rods. At long times, in the isotropic phase, both quantities should superpose in the diffusive regime and the diffusion coefficient $D$ can be calculated from the slope of the MSD where $g_{0}(t)=6Dt$.

Both quantities are shown in Fig \ref{fig:diffcoeff} for ungrafted rods in the isotropic phase at a concentration of $\phi=0.1$. The linear fit (pink line) to the plateau seen at long times yields a diffusion coefficient of $D=4.1\times10^{-3}$ $(\sigma^{2}/\tau)$. From this a reference timescale of an ungrafted rod to diffuse its own length ($\sim 30 \sigma$) may be approximated as $\tau_{\rm D} = \sigma^{2}/D \approx 7k\tau$. 

\end{document}